\documentclass[aps,prd,nofootinbib]{revtex4-2}
\usepackage[final]{graphicx}
\makeatletter\Gin@draftfalse\makeatother
\usepackage{amsmath,amssymb,epsfig}
\usepackage{paralist}
\usepackage{comment}
\usepackage{wrapfig}
\usepackage{url}
\usepackage[colorlinks=true,linkcolor=blue,citecolor=blue,urlcolor=blue]{hyperref}

\renewcommand{\vec}[1]{\boldsymbol{\mathrm{#1}}}

\begin{document}

\title{Gravitational lensing for interstellar power transmission}

\author{Slava G. Turyshev}

\affiliation{
Jet Propulsion Laboratory, California Institute of Technology,\\
4800 Oak Grove Drive, Pasadena, CA 91109-0899, USA}

\date{\today}

\begin{abstract}
 
We investigate the propagation of monochromatic electromagnetic waves through multiple, well-separated, monopole gravitational lenses in the thin-lens/eikonal approximation.  For the axially aligned transmitter--lens(es)--receiver geometry, the relevant diffraction integrals can be evaluated analytically, yielding closed-form expressions for the point-spread function (PSF) and for the aperture-averaged gain.  A single gravitational lens can enhance transmission when it is used either at the transmitting end or at the receiving end.  For a two-lens link, the second lens focuses the signal into a much smaller diffraction pattern; however, for optical wavelengths and metre-scale receiving apertures this fine structure is aperture-averaged, and the averaged additional gain becomes independent of the second lens mass.  We estimate photon rates and shot-noise-limited SNRs for these idealized configurations.  The results indicate that gravitational lensing can in principle support high-SNR interstellar optical power links, subject to stringent alignment, finite-aperture, transmitter-beam, and coronal-background assumptions.

\end{abstract}

\maketitle

\section{Introduction}
\label{sec:introduction}

Recently, we have explored the optical properties of the solar gravitational lens (SGL) \cite{Turyshev:2017,Turyshev-Toth:2017} and have shown that the SGL is characterized by a significant light amplification and angular resolution. As such, the SGL provides unique  capabilities for direct imaging and spectroscopy of faint targets such as exoplanets in our stellar neighborhood \cite{Turyshev-Toth:2022a,Turyshev-Toth:2022b}.  

It can be assumed that pairs of stellar gravitational lenses could facilitate energy transmission across interstellar distances, utilizing equipment of a scale and power akin to that currently employed for interplanetary communications \cite{vonEshleman:1979}. There is a prevailing expectation that such a configuration would benefit from the light amplification by both lenses, thus, enabling significant increases in the signal-to-noise ratio (SNR) of a received signal \cite{Maccone-book:2009,Maccone:2011,Clark:2018,Kerby-Wright:2021,Tusay:2022,Gillon:2022}. However, a comprehensive analysis of these transmission scenarios remains to be undertaken. 

Our objective here is to examine light propagation in the case of multi-lens systems and to evaluate the associated light amplification for various practical transmission scenarios. To do that, we will rely on the wave-theoretical treatment of the gravitational lensing phenomena and will use analytical tools that we developed in our prior studies of spherically-symmetric gravitational lenses  \cite{Turyshev-Toth:2017,Turyshev-Toth:2022a,Turyshev-Toth:2022b,Turyshev-Toth:2021-multipoles,Turyshev-Toth:2021-all-regions,Turyshev-Toth:2020-im-extend}. As such, these methods can be seamlessly adapted to explore power transmission within the multi-lens configurations. 

In this study, our objective is to systematically explore the optimal transmission scenarios where the alignment of the light source, optical lenses, and receiver is either precise or approximately so. This alignment is crucial for maximizing efficiency in photon transmission. By focusing on such configurations, we aim to identify and analyze transmission architectures that are not only theoretically efficient but also practically implementable with existing or emerging optical technologies. This approach allows for the investigation of feasible solutions in the realm of advanced photonics and optical engineering.

This paper is organized as follows: In Section~\ref{sec:wave}, we present the wave-theoretical tools to describe the propagation of EM waves in a gravitational field. In Section \ref{sec:light-amp}, we consider various lensing geometries that involve one and two gravitational lenses, and derive the relevant light amplification factors.  In Section~\ref{sec:power-X}, we discuss power transmission with lensing configurations utilizing both a single lens and a pair of lenses. In Section~\ref{sec:SNR} we evaluate detection sensitivity in various cases considered and evaluate the relevant SNRs. Our conclusions are presented in Section~\ref{sec:summary}.
To streamline the discussion, we moved some material to 
Appendices. Appendix \ref{sec:alt-der} presents an alternative evaluation of the diffraction integral in the case of a two-lens transmission. Appendix \ref{sec:path-int} presents a path integral formulation.

\section{EM waves in a gravitational field}
\label{sec:wave}

As an electromagnetic (EM) wave travels in the vicinity of a stellar gravitational lens, its interaction with the lens' gravity causes the wave to scatter and diffract.  In Refs.~\cite{Turyshev-Toth:2017,Turyshev-Toth:2021-multipoles,Turyshev-Toth:2021-all-regions}, while studying the Maxwell equations on a weak-gravity spacetime background, we developed a solution to the Mie problem for the diffraction of the EM waves in the presence of a large gravitating body (see also \cite{Herlt-Stephani:1976,Born-Wolf:1999})
and found the EM field at an image plane located in the focal region of the lens. Here we briefly summarize the relevant results.

\subsection{Diffraction of light in a gravitational field}
\label{sec:basics}

\begin{figure}
\vspace{-10pt}
  \begin{center}
\includegraphics[width=0.55\linewidth]{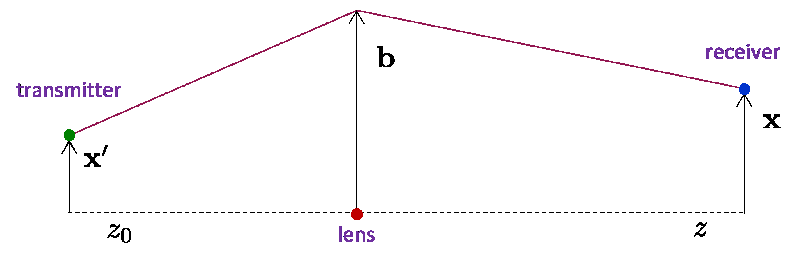}
  \end{center}
  \vspace{-15pt}
  \caption{A lens-centric  geometry for power transmission via gravitational lensing showing the transmitter, the lens, and the receiver. Also shown is the distance from the lens to the transmitter plane, $z_0$, and that from the lens to the receiver plane, $z$.  
  }
\label{fig:geom}
\end{figure}

We consider a stellar gravitational lens with Schwarzschild radius $r_g=2GM/c^2$, where $M$ is the lens mass.  We assume that the light originates at a finite but large distance from the lens.  We introduce a lens-centric Cartesian coordinate system (Fig.~\ref{fig:geom} shows overall geometry of the gravitational lensing system) with its $z$-axis oriented along the unperturbed direction of propagation of the incident wave, given by the vector $\vec k$.  We also introduce the light ray's impact parameter, $\vec b$, with $(\vec b \cdot \vec k)=0$. In this coordinate system, the source is positioned at $(\vec x',-z_0)$, while the receiver is on the image plane with coordinates $(\vec x,z)$, located in the lens' strong interference region at a distance $z\geq b^2/2r_g$ from the lens, with $b=|\vec b|$. 
Finally, introducing cylindrical coordinate system  $(\rho,\phi,z)$  centered at the lens,  the quantities mentioned above may be given as follows
{}
\begin{align}
\vec k&=(0,0,1), \qquad 
{\vec b}=b(\cos\phi_\xi,\sin \phi_\xi,0), \qquad 
{\vec x}'=\rho'(\cos\phi',\sin \phi',0), \qquad 
{\vec x}=\rho(\cos\phi,\sin \phi,0).
\label{eq:note-x}
\end{align}

With the definitions in Eq.~(\ref{eq:note-x}), the EM field on an image plane takes the following form (see details in  \cite{Turyshev-Toth:2021-multipoles,Turyshev-Toth:2021-all-regions}):
{}
\begin{eqnarray}
    \left( \begin{aligned}
{E}_\rho& \\
{H}_\rho& \\
  \end{aligned} \right) =\left( \begin{aligned}
{H}_\phi& \\
-{E}_\phi& \\
  \end{aligned} \right) &=&
A(\vec x', \vec x)e^{-i\omega t}
 \left( \begin{aligned}
 \cos\phi& \\
 \sin\phi& \\
  \end{aligned} \right)+{\cal O}\Big(r_g^2,\rho^2/z^2\Big),
  \label{eq:DB-sol-rho}
\end{eqnarray}
with the remaining components being small, i.e., $({E}_z, {H}_z)\propto {\cal O}({\rho}/{z})$. 
 
Assuming the validity of eikonal and the thin lens approximation, the Fresnel-Kirchhoff diffraction formula \cite{Landau-Lifshitz:1988,Born-Wolf:1999} yields the following expression for the wave's amplitude at the observer (receiver) location
 {}
\begin{eqnarray}
A(\vec x', \vec x) &=&E_0
 \frac{k}{i z_0z}\frac{1}{2\pi}\iint d^2\vec b \,\hat A_0(\vec b)e^{ik S(\vec x', \vec b, \vec x)},
  \label{eq:amp-FK}
\end{eqnarray}
where $k=2\pi/\lambda$ is the wavenumber and  $S(\vec x', \vec x, \vec b)$  is the effective path length (eikonal) along a path from the source positioned at $(\vec x', -z_0)$ to the observer's position at $(\vec x,z)$ via a point $(\vec b,0)$ on the lens plane  (see Fig.~\ref{fig:geom})
 {}
\begin{eqnarray}
S(\vec x', \vec b, \vec x) &=&
 \sqrt{(\vec b-\vec x')^2+z_0^2}+ \sqrt{(\vec b-\vec x)^2+z^2}-\psi(\vec b)= \nonumber\\
 &=&z_0+z+\frac{(\vec x-\vec x')^2}{2(z_0+z)}+\frac{z_0+z}{2z_0z}\Big(\vec b-\frac{z_0}{z_0+z}\vec x-\frac{z}{z_0+z}\vec x'\Big)^2-\psi(\vec b)+{\cal O}\Big(\frac{b^4}{z_0^3},\frac{b^4}{z^3}\Big),
  \label{eq:amp-S}
\end{eqnarray}
where the last term in this expression is the gravitational phase shift, $\psi(\vec b) $, that is acquired by the EM wave as it propagates along its geodetic path from the source to the image plane on the background of the gravitational field with potential, $U$,  that has the form (see discussion in \cite{Turyshev-Toth:2021-multipoles,Turyshev-Toth:2022-STF}):
{}
\begin{eqnarray}
\psi(\vec b)&=&-r_g\ln(4k^2zz_0)+2r_g\ln(k b)+{\cal O}(r_gJ_n),
  \label{eq:gp}
\end{eqnarray}
where the sign convention follows from using the Newtonian potential $U<0$ in Eq.~(\ref{eq:amp-S}).  The coefficients $J_n$, $n\in\{2,3,4,\ldots\}$, are the spherical-harmonic coefficients representing the mass distribution inside the stellar lens \cite{Turyshev-Toth:2021-multipoles}. Using the monopole phase convention (\ref{eq:gp}), the wave amplitude on the observer plane can be written as
{}
\begin{eqnarray}
A(\vec x', \vec x)&=& A_0(\vec x', \vec x)F(\vec x', \vec x),
  \label{eq:amp00}
\end{eqnarray}
where $A_0(\vec x', \vec x)$ is the wave amplitude at the receiver (observer) in the absence of the gravitational potential $U$:
{}
\begin{eqnarray}
A_0(\vec x', \vec x)&=& \frac{E_0}{z_0+z}e^{ikS_0(\vec x',\vec x)},
\qquad S_0(\vec x',\vec x)=z_0+z+\frac{(\vec x-\vec x')^2}{2(z_0+z)},
  \label{eq:amp}
\end{eqnarray}
with $S_0(\vec x',\vec x)$ being the path length along a straight path from $\vec x'$ to $\vec x$. In the case of a monopole lens, the amplification factor $F(\vec x', \vec x)$ is given by the following form of a diffraction integral
{}
\begin{eqnarray}
F(\vec x',\vec x) &=&\frac{z_0+z}{z_0z} \frac{ke^{i\phi_{\tt G}}}{2\pi i}\iint d^2\vec b \,\hat A_0(\vec b)e^{ikS_1(\vec x', \vec b, \vec x)},\nonumber\\
 S_1(\vec x',\vec b, \vec x)&=&\frac{z_0+z}{2z_0z}\Big(\vec b-\frac{z_0}{z_0+z}\Big(\vec x+\frac{z}{z_0}\vec x'\Big)\Big)^2-2r_g\ln(k b),~~~~
  \label{eq:amp-Am+}
\end{eqnarray}
where the phase factor is given as $\phi_{\tt G}=kr_g\ln(4k^2zz_0)$ and  $S_1(\vec x',\vec b, \vec x)$  is the Fermat potential along a path from the source position $\vec x'$ to the observer position $\vec x$ via a point $\vec b$ on the lens plane. $\hat A_0(\vec b)$ is the surface brightness of the extended source. The first term in $S_1(\vec x',\vec b,\vec x)$  is the difference of the geometric time delay between a straight path from the source to the observer and a deflected path. The second term is due to the time delay in the gravitational potential of the lens object (i.e., the Shapiro time delay).

In the case of an isolated spherically-symmetric gravitational lens positioned on the optical axis, $\vec x'=0$, collecting all the relevant terms, we present the amplification factor on the receiver (observer) plane 
{}
\begin{eqnarray}
F(\vec x) &=&
\frac{z_0+z}{z_0z}\frac{ke^{i\phi_{\tt G}}}{2\pi i}
 \iint d^2\vec b \,\hat A_0(\vec b)\exp\Big[ik\Big(\frac{z_0+z}{2z_0z}\Big(\vec b-\frac{z_0}{z_0+z}\vec x\Big)^2-2r_g\ln(k b)\Big)\Big].
  \label{eq:amp-Am}
\end{eqnarray}

We consider the case of two non-interacting thin gravitational lenses.
We denote $z_{\tt t}$ to be the transmitter's distance from the first lens, $z_{12}$ is the distance between the lenses, and $z_{\tt r}$ is the distance between the second lens and the receiver.  We assume that the lenses are at a very large distance, $z_{12}\gg z_{\tt t}, z_{\tt r}$, from each other; also $z_{\tt t}$ and $z_{\tt r}$ are in the focal regions of the respective lenses. This allows us to treat the light propagation independently for each lens. 

Furthermore, as we are interested to evaluate the largest light amplification, we  consider the most favorable transmission geometry: we assume that all participants -- the source (or transmitter), the first and the second monopole lenses, and the observer (or receiver) -- are all situated on the same line -- the primary optical axis of the lensing system. Clearly, any deviation from this axially-symmetric geometry will reduce the energy transmitting efficiency.

In this scheme, the light emitted by the source is diffracted by the gravitational field of the first lens, that focuses and amplifies light which now  becomes the source for the second lens.  This EM field encounters the second lens, then the third lens and so on and ultimately it reaches the observer that is positioned in the focal region of the last lens.  

\subsection{Transmission geometry}
\label{sec:transmit-geom}

A fully self-consistent wave-optical solution for a finite-aperture transmitter at finite distance, emitting a prescribed spherical or Gaussian beam toward a gravitational lens, is not developed here.  Instead, we use the known monopole-lens diffraction solution for incident waves that may be approximated locally as plane waves, and we combine it with diffraction-limited transmitter-beam estimates in the link-budget sections below.

It is known that when an EM wave, originating at infinity, travels near a gravitational body, its wavefront experiences bending. In general relativity, this deflection angle is $\theta_{\tt gr}={2r_g}/{b}$. As a result, a massive body acts as a lens by focusing the EM radiation (i.e., the light rays intersect the optical axis) at the distance, $z$, that is determined as
{}
\begin{eqnarray}
\frac{b}{z}=\theta_{\tt gr}
\qquad \Rightarrow \qquad z= \frac{b^2}{2r_g}. 
\label{eq:foc_z}
\end{eqnarray}

In \cite{Turyshev-Toth:2019-extend}, we have shown that  expression for $z$ from (\ref{eq:foc_z}) is modified when the light rays are coming from a source located at a finite distance, $z_0$, from the lens. In this case, defining $\alpha=b/z_{0}$ and using small angle approximation, the  new expression reads
{}
\begin{eqnarray}
\frac{b}{z}=\theta_{\tt gr}-\alpha
\qquad \Rightarrow \qquad z= \frac{b^2}{2r_g}\frac{1}{1-b^2/2r_gz_0} . 
\label{eq:foc_z2}
\end{eqnarray}

Clearly, the transmitter's distance with respect to the lens, determines four signal transmission regimes, namely
\begin{enumerate}[i).]
\item For $\theta_{\tt gr}<\alpha$ or when $0\leq z_0< {b^2}/{2r_g}$ there is no transmission. In this case, the light rays either are completely absorbed by the lens or are not able to focus, e.g., do not reach the optical axis on the other side behind the lens. 

\item For $\theta_{\tt gr}=\alpha$ or when $z_0= {b^2}/{2r_g}$, the focal distance $ z$ is infinite, implying that after passing by the lens the light rays are collimated, never reaching the optical axis. In this case, there will be shadow behind the lens except for the presence of the bright spot of Arago that weakens with distance from the lens \cite{Turyshev-Toth:2018,Turyshev-Toth:2018-grav-shadow}.

\item For $\theta_{\tt gr}>\alpha$ or when $z_0> {b^2}/{2r_g}$, the light will begin to focus at a finite distance from the lens, forming all the regions relevant for the diffraction problem including the shadow,  interference region, and that of the geometric optics (see Fig.~4 in \cite{Turyshev-Toth:2017}).  According to  (\ref{eq:foc_z2}),  these regions will be formed farther from the lens, compared to  (\ref{eq:foc_z}).

\item For $\theta_{\tt gr} \gg \alpha$  or when $z_0\gg {b^2}/{2r_g}$, the point source is effectively at infinity and, depending on the impact parameter, the light rays will form all the typical diffraction regions, beginning at the distance given by (\ref{eq:foc_z}). 

\end{enumerate}

As a result, for optimal reception, an observer needs to be positioned in the focal region of a lens at the distance given by (\ref{eq:foc_z2}). Same logic works when a transmission link includes another lens which is then followed by a receiver.

\section{Light amplification in various transmission scenarios}
\label{sec:light-amp}

We are now in a position to explore the light propagation in the gravitational field and to describe light amplification in various lensing configurations, including scenarios with one or two linearly-aligned gravitational lenses.

\subsection{One lens transmission} 
\label{sec:single-lens}

We consider a stellar monopole gravitational lens with mass $M_1$, Schwarzschild radius $r_{g_1}=2GM_1/c^2$, and physical radius $R_1$. We assume that the transmitter is positioned in the lens' focal region and placed on the optical axis, defined as the line connecting the transmitter, the lens, and the receiver. Thus, in (\ref{eq:amp00})--(\ref{eq:amp-Am}) we can set $\vec x'=0$. 

Note that there are two distinct architectures to form a transmission link that are determined by where the transmitter and receiver are placed with respect to the lens: i). In one case, the transmitter is positioned close to the lens but at the distance larger than the beginning of the focal region, namely $z_{\tt t}\geq R^2_1/2r_{g_1}=547.8\, (R_1/R_\odot)^2(M_\odot/M_1)~{\rm AU}$.  (Note that in the case of the Sun, the solar corona increases the distance from where practical transmissions may occur. Thus, for $\lambda\sim 1~\mu$m,  these ranges are beyond $\sim 650$~AU from the Sun.)  In this case, the receiver is placed at an interstellar distance from the lens, so that $z_{\tt r}\gg z_{\tt t}$. ii). In another case, the positions are switched, so that the transmitter is now placed at an interstellar distance from the lens, so that $z_{\tt t}\gg R^2_1/2r_{g_1}$, but the receiver is at the lens' focal region, $z_{\tt r}\geq  R^2_1/2r_{g_1}$. Our formulation below will cover both of these cases. 

Considering diffraction on a single lens, with the help of  (\ref{eq:amp-Am}) and taking $\hat A_0(\vec b)=1$, we determine the amplification factor of the EM wave at the observer's plane that is positioned a distance $z_{\tt r}$ from the lens  as below
{}
\begin{eqnarray}
F_{\tt 1GL}(\vec x) &=&
 \frac{ke^{i\phi_{\tt G1}}}{i \tilde z_1}\frac{1}{2\pi}
 \iint d^2\vec b_1\exp\Big[ik\Big(
 \frac{1}{2\tilde z_1}\Big(\vec b_1-\frac{\tilde z_{1}}{z_{\tt r}}\vec x\Big)^2-2r_{g_1}\ln(k b_1)\Big)\Big],
  \qquad {\rm where}\qquad 
 \tilde z_1=\frac{z_{\tt t}z_{\tt r}}{z_{\tt t}+z_{\tt r}},
  \label{eq:amp-A1*}
\end{eqnarray}
where $\phi_{\tt G1}=kr_{g_1}\ln(4k^2z_{\tt t}z_{\tt r})$. After re-arranging the terms and removing the spherical wave \cite{Wolf-Gabor:1959,Born-Wolf:1999} (which represents the action of a convex lens that transforms incident plane waves to spherical waves, focusing at the focal point, see discussion in \cite{Turyshev-Toth:2020-im-extend}), we present (\ref{eq:amp-A1*}) as
{}
\begin{eqnarray}
F_{\tt 1GL}(\vec x) &=&
 \frac{ke^{i\phi_{\tt G1}}}{i \tilde z_1}\frac{1}{2\pi}\int_0^\infty b_1db_1 \exp\Big[ik\Big(\frac{b^2_1}{2 \tilde z_{1}} -2 r_{g_1}\ln(k b_1)\Big)\Big]\int_0^{2\pi} d\phi_{\xi_1}\exp\Big[-i\Big(k\frac{b_1\rho}{z_{\tt r}} \cos[\phi_{\xi_1}-\phi]\Big)\Big]=\nonumber\\
 &=&
  \frac{ke^{i\phi_{\tt G1}}}{i \tilde z_1}
\int_0^\infty b_1db_1 J_0\Big(k\frac{b_1\rho}{z_{\tt r}}\Big)\exp\Big[ik\Big(\frac{b^2_1}{2 \tilde z_{1}} -2 r_{g_1}\ln(k b_1)\Big)\Big].
  \label{eq:amp-A1*23}
\end{eqnarray}

Considering the four scenarios detailed in Sec.~\ref{sec:transmit-geom}: Case i) is trivial and does not require a formal treatment.  Case ii) corresponds to $\theta_{\tt gr}=\alpha$, or equivalently $z_{\tt t}=b_1^2/(2r_{g_1})$.  In this limiting case the rays emerging behind the lens are collimated rather than focused at a finite distance.  The field has the character of an Arago spot for an occulting disk; using the stellar radius $R_1$ as the effective edge, one obtains
{}
\begin{eqnarray}
F^0_{\tt 1GL}(\vec x)&=&\Big(1+\frac{z_{\tt r}}{z_{\tt t}}\Big)
e^{i\big(\phi_{\tt G1}+\frac{kR^2_1}{2  z_{\tt r}}\big)} J_0\Big(k\frac{R_1\rho}{z_{\tt r}}\Big) \qquad \Rightarrow \qquad
A^0_{\tt 1GL}(\vec x)=\frac{E_0}{z_{\tt t}}
e^{i\big(S_0(\vec x)+\phi_{\tt G1}+\frac{kR^2_1}{2  z_{\tt r}}\big)} J_0\Big(k\frac{R_1\rho}{z_{\tt r}}\Big),
  \label{eq:amp-A1*230}
\end{eqnarray}
where $S_0(\vec x )=z_{\tt t}+z_{\tt r}+\rho^2/2(z_{\tt t}+z_{\tt r})$. Result (\ref{eq:amp-A1*230}) has the properties of the bright spot of Arago \cite{Harvey-Forgham:1984}. Note that the stellar atmospheres will make the edges of the spherical lenses optically softer, thus severely affecting formation of the spot to the point of completely washing it out. As a result, this scenario is not very useful for power transmission.

Next, we examine the scenarios, Case iii) and Case iv), as highlighted in Sec.~\ref{sec:transmit-geom}. These cases are similar and can be characterized in the same manner.  For that, we take the last integral in (\ref{eq:amp-A1*23}) with the method of stationary phase \cite{Turyshev-Toth:2017} to determine the impact parameter for which the phase is stationary:
{}
\begin{eqnarray}
b_1=\sqrt{2r_{g_1} \tilde z_{1}},
  \label{eq:impar}
\end{eqnarray} 
Substituting (\ref{eq:impar}) yields the following result for the amplification factor:
{}
\begin{eqnarray}
F_{\tt 1GL}(\vec x)
&=&\sqrt{2\pi k r_{g_1}}  e^{i\varphi_{1}}
J_0\Big(k\frac{\sqrt{2r_{g_1} \tilde z_1}}{z_{\tt r}}\rho \Big),
  \label{eq:amp-A1*24}
\end{eqnarray}
where  $\varphi_{1}$ is given as $\varphi_{1}=\phi_{\tt G1}+k\big(r_{g_1}-r_{g_1}\ln(k r_{g_1})-r_{g_1}\ln(2k \tilde z_{1})
\big)-{\textstyle\frac{1}{4}}\pi.$

To evaluate the light amplification of a single lens, we first determine its point-spread function (PSF). For that, we use the generic solution for the EM field (\ref{eq:DB-sol-rho}) with solution (\ref{eq:amp}) together with (\ref{eq:amp-A1*24}), and study the Poynting vector, ${\vec S}=(c/4\pi)\big<\overline{[{\rm Re}{\vec E}\times{\rm Re}{\vec H}]}\big>$, that describes the energy flux in the image plane \cite{Born-Wolf:1999}. Normalizing this flux to the time-averaged value that would be observed if the gravitational field of the first lens were absent, $|\overline{\vec S}_0|=(c/8\pi)E_0^2/(z_{\tt t}+z_{\tt r})^2$, we determine the PSF of a single lens ${\tt PSF}_{\tt 1GL}=|{\vec S}|/|\overline{\vec S}_0|$ as below:
{}
\begin{align}
{\tt PSF}_{\tt 1GL}({\vec x})&
=2\pi k r_{g_1}  J^2_0\Big(k\frac{\sqrt{2r_{g_1} \tilde z_1}}{ z_{\tt r}}\rho \Big).
\label{eq:S_z*6z-mu2}
\end{align}
Therefore, the largest value for the light amplification factor of a single lens is realized when $\rho=0$, yielding \cite{Turyshev:2017,Turyshev-Toth:2017}
{}
\begin{eqnarray}
\mu^0_{\tt 1GL} &=&
2\pi k r_{g_1} \simeq 1.17\times 10^{11} \Big(\frac{M_1}{M_\odot}\Big)\Big(\frac{1\,\mu{\rm m}}{\lambda}\Big).
  \label{eq:amp-PSF1}
\end{eqnarray}

\begin{wrapfigure}{R}{0.43\textwidth}
\vspace{-10pt}
  \begin{center}
   \rotatebox{90}{\hskip 35pt  {\small Normalized PSF}}
\hskip 0pt
\includegraphics[width=0.95\linewidth]{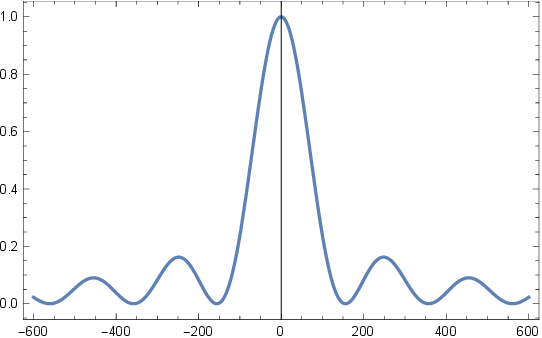}
 \rotatebox{0}{\hskip 30pt  {\small Distance from the optical axis, $\rho$,  [m]}}
  \end{center}
  \vspace{-15pt}
  \caption{Normalized PSF of a single lens in the transmitting lens scenario (\ref{eq:S_z*6z-mu2+}), for the relevant physical parameters given in (\ref{eq:theta1-t}). Note that the first zero here is much farther out from the optical axis.
  }
\label{fig:psf10}
\vspace{-10pt}
\end{wrapfigure}

Nominally, amplification (\ref{eq:amp-PSF1}) is realized for a telescope with the diameter that is less than the characteristic pattern of the Bessel function $J_0(k\rho \sqrt{{2r_{g_1} \tilde z_{1}}}/z_{\tt r})$. 
In practice, if the telescope diameter $d$ is  larger than that of the first zero of the projected Airy pattern (\ref{eq:S_z*6z-mu2}), it would average the light amplification over that aperture (see discussion in \cite{Turyshev-Toth:2017}.) We shall discuss this fact when we look at different transmission scenarios below.

Coming back to the transmission scenarios iii) and iv) discussed in Sec.~\ref{sec:transmit-geom}, we note that both are described by the results above.  In scenario iii), the transmitter is near the lens and the receiver is far away, $z_{\tt t}\ll z_{\tt r}$.  In scenario iv), the transmitter is far from the lens and the receiver is in the focal region, $z_{\tt t}\gg z_{\tt r}$. 

With solution (\ref{eq:S_z*6z-mu2}), we may now consider the two cases with drastically different positions of the transmitter and the receiver that were discussed at the beginning of this section.  

\subsubsection{Transmitting lens scenario}
\label{sec:x-dom}

In the case, when $z_{\tt t}\ll z_{\tt r}$,  the effective distance $\tilde z_1$ from (\ref{eq:amp-A1*}) reduces to $\tilde z_1={z_{\tt t}z_{\tt r}}/(z_{\tt t}+z_{\tt r}) \simeq  z_{\tt t}$. Therefore, the PSF of the transmission with a single lens  (\ref{eq:S_z*6z-mu2}) takes the form
{}
\begin{align}
{\tt PSF}^{\tt t}_{\tt 1GL}({\vec x})&
\simeq2\pi k r_{g_1}  J^2_0\Big(k\frac{\sqrt{2r_{g_1} z_{\tt t}}}{ z_{\tt r}}\rho \Big).
\label{eq:S_z*6z-mu2+}
\end{align}
We shall call this case the transmitting lens scenario, thus there will be superscript $\{\}^{\tt t}$ on the relevant quantities.

Note that in this case, an observer in the focal region of the lens, positioned on the optical axis at $z_{\tt r}\gg z_{\tt t}$ from it, will see an Einstein ring around the lens with the radius $\theta^{\tt t}_1$ given as (see details in \cite{Turyshev-Toth:2023}) below:
{}
\begin{eqnarray}
\theta^{\tt t}_1=\frac{\sqrt{2r_{g_1} z_{\tt t}}}{z_{\tt r}}
 \simeq 2.46 \times 10^{-9}\,\Big(\frac{M_1}{M_\odot}\Big)^\frac{1}{2}\Big(\frac{z_{\tt t}}{650\,{\rm AU}}\Big)^\frac{1}{2}\Big(\frac{10\,{\rm pc}}{z_{\tt r}}\Big)~ {\rm rad}.
  \label{eq:theta1-t}
\end{eqnarray}

We observe that the first zero of the projected Airy pattern in (\ref{eq:S_z*6z-mu2+}) occurs at the distance of $\rho^{\tt t}_{\tt 1GL}=2.40483 \,/(k\theta^{\tt t}_1)\simeq 155.82\,{\rm m}\,({\lambda}/{1~\mu{\rm m}})({M_\odot}/{M_1})^\frac{1}{2}(650~{\rm AU}/z_{\tt t})^\frac{1}{2}(z_{\tt r}/10\,{\rm pc})$ from the optical axis, which is large and aperture averaging may not be important. Fig.~\ref{fig:psf10} shows the relevant behavior of this PSF. Therefore, while considering the relevant transmission links, one may have to use the entire PSF from (\ref{eq:S_z*6z-mu2+}) with its maximal value $\mu^0_{\tt 1GL}$ given by (\ref{eq:amp-PSF1}).

\subsubsection{Receiving lens scenario}
\label{sec:receiving-lens}

In the case, when $z_{\tt t}\gg z_{\tt r}$,  the effective distance $\tilde z_1$ from (\ref{eq:amp-A1*}) reduces to $\tilde z_1={z_{\tt t}z_{\tt r}}/(z_{\tt t}+z_{\tt r}) \simeq  z_{\tt r}$. Therefore, the PSF of a single lens transmission (\ref{eq:S_z*6z-mu2}) takes the form
{}
\begin{align}
{\tt PSF}^{\tt r}_{\tt 1GL}({\vec x})&
\simeq2\pi k r_{g_1}  J^2_0\Big(k\sqrt{\frac{2r_{g_1}}{z_{\tt r}}}\rho \Big).
\label{eq:S_z*6z-mudd+}
\end{align}
We shall call this case the receiving lens scenario, thus there will be superscript $\{\}^{\tt r}$ on the relevant quantities.

Note that in this case, a receiver in the focal region of the lens, with the transmitter at $z_{\tt t}\gg z_{\tt r}$, will see an Einstein ring around the lens with the angular radius $\theta^{\tt r}_1$ given as (see details in \cite{Turyshev-Toth:2023}):
{}
\begin{eqnarray}
\theta^{\tt r}_1=\sqrt{\frac{2r_{g_1}}{z_{\tt r}}}
 \simeq 7.80 \times 10^{-6}\,\Big(\frac{M_1}{M_\odot}\Big)^\frac{1}{2}\Big(\frac{650\,{\rm AU}}{z_{\tt r}}\Big)^\frac{1}{2}~ {\rm rad},
  \label{eq:theta1-r}
\end{eqnarray}
which is much larger than that obtained for the transmitting lens scenario (\ref{eq:theta1-t}). 

We observe that, in this case the first zero of the projected Airy pattern (\ref{eq:S_z*6z-mudd+}) occurs at $\rho^{\tt r}_{\tt 1GL}=2.40483/(k\theta^{\tt r}_1)\simeq 4.91\,{\rm cm}\,({\lambda}/{1~\mu{\rm m}})({M_\odot}/{M_1})^\frac{1}{2}(z_{\tt r}/650~{\rm AU})^\frac{1}{2}$, which is very small and needs to be aperture-averaged. Fig.~\ref{fig:psf-12}  demonstrates the relevant behavior of the PSF (\ref{eq:S_z*6z-mudd+}) at very short spatial scales of a few cm. To estimate the impact of the large aperture on the light amplification, we average the result (\ref{eq:S_z*6z-mudd+}) over the aperture of the telescope and, using approximation for the Bessel functions for large arguments \cite{Abramovitz-Stegun:1965}, we determine:
{}
\begin{eqnarray}
\left<\mu^{\tt r}_{\tt 1GL}\right>&=&
\frac{1}{\pi ({\textstyle\frac{1}{2}}d_{\tt r})^2}
\int_0^{{\textstyle\frac{1}{2}}d_{\tt r}}\int_0^{2\pi}
{\tt PSF}^{\tt r}_{\tt 1GL}({\vec x})\,\rho d\rho d\phi
\nonumber\\
&=&
2\pi k r_{g_1}
\Big[
J^2_0\big(k \theta^{\tt r}_1{\textstyle\frac{1}{2}}d_{\tt r}\big)
+
J^2_1\big(k \theta^{\tt r}_1{\textstyle\frac{1}{2}}d_{\tt r}\big)
\Big]
\simeq
\frac{4 \sqrt{2r_{g_1} z_{\tt r}}}{d_{\tt r}}
\nonumber\\
&=&
3.03\times 10^{9}
\Big(\frac{M_1}{M_\odot}\Big)^\frac{1}{2}
\Big(\frac{z_{\tt r}}{650\, {\rm AU}}\Big)^\frac{1}{2}
\Big(\frac{1~{\rm m}}{d_{\tt r}}\Big),
\label{eq:mu_av}
\end{eqnarray}
where we recognize the well-known result  obtained in \cite{Turyshev-Toth:2017}.
Compared to (\ref{eq:amp-PSF1}), the aperture-averaging given by (\ref{eq:mu_av}) leads to a reduction in the light amplification by a factor of $\simeq 38.46$, thus resulting only in 2.6\% of the maximal value suggested by (\ref{eq:amp-PSF1}). Nevertheless, the overall result is still rather impressive. 

The extension to two aligned lenses is developed next in Sec.~\ref{sec:2lenses}.

\subsection{Two lens transmission} 
\label{sec:2lenses}
    
\begin{figure}[t]
  \begin{center}
   \rotatebox{90}{\hskip 35pt {\small Normalized PSF}}
   \hskip 0pt
   \includegraphics[width=0.55\linewidth]{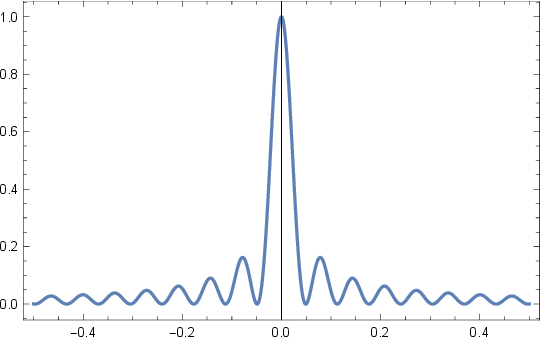}

   \vspace{-4pt}
   {\small Distance from the optical axis, $\rho$, [m]}
  \end{center}
  \vspace{-10pt}
  \caption{Normalized PSFs for a single lens in the receiving-lens scenario
  (\ref{eq:S_z*6z-mudd+}), with $M_1=M_\odot$ and the other physical
  parameters from (\ref{eq:theta1-r}), and for a double-lens transmission
  (\ref{eq:amp-psf2-av=+}), with $M_2=M_\odot$ and the relevant parameters
  from (\ref{eq:theta1-r=}).  The normalized profiles are identical for
  these fiducial parameters; their absolute normalizations are given by
  Eqs.~(\ref{eq:mu_av}) and (\ref{eq:mu_av=}).}
\label{fig:psf-12}
\end{figure}

We consider two lenses, positioned on the same line -- the optical axis -- and separated from each other by the distance $z_{12}$. The first lens has the parameters used in Sec.~\ref{sec:single-lens}, while lens 2 is given by the mass, $M_2$, Schwarzschild radius $r_{g_2}=2GM_2/c^2$, and radius $R_2$. To describe power transmission via a two-lens system, one can develop a scheme similar to one in Fig.~\ref{fig:geom}. Next, we extended the expression for the optical path (\ref{eq:amp-S}) by modeling the contribution of the second lens to the overall optical path, including that from its gravitational field (see also discussion in Appendix~\ref{sec:path-int}.) 

As a result, we treat the EM wave that already passed by lens 1 to be the source of light incident on lens 2.  For an observer at the lens 1 plane, the angle subtended by the physical radius of lens 2 is small $R_2/z_{12}\simeq \theta_1$, where $\theta_1$ is the angle that determines the diffraction pattern of the incident field  (\ref{eq:amp-A1*24}). This angle is explicitly evaluated on the lens 2 plane, so that distance $z_{\tt r}$ in (\ref{eq:amp-A1*24})  is replaced by $z_{12}$, yielding the expression that is relevant for the transmission scenario discussed here  
{}
\begin{eqnarray}
\theta_{1}=\frac{\sqrt{2r_{g_1} \tilde z_1}}{z_{12}}
 \simeq 2.46 \times 10^{-9}\,\Big(\frac{M_1}{M_\odot}\Big)^\frac{1}{2}\Big(\frac{z_{\tt t}}{650\,{\rm AU}}\Big)^\frac{1}{2}\Big(\frac{10\,{\rm pc}}{z_{12}}\Big)~ {\rm rad}.
  \label{eq:theta1-t12}
\end{eqnarray}
Therefore, as $R_2/z_{12}\simeq \theta_1$, this lensing geometry corresponds to the strong interference regime of  diffraction on lens 1 (see \cite{Turyshev-Toth:2017} for description). The analytical description of this process was already addressed in Sec.~\ref{sec:single-lens}. 

In the two-lens calculation, the inner diffraction integral gives the one-lens amplification at the plane of lens 2.  The free quadratic propagation phase from the transmitter to that plane is kept in the outer propagation kernel below.  Thus we write
{}
\begin{eqnarray}
{\cal F}_{\tt 1GL}(\vec b_2)
&=&\sqrt{2\pi k r_{g_1}}\,
J_0\big(k \theta_{1} b_2 \big),
\label{eq:amp-A1*25}
\end{eqnarray}
with constant phases absorbed into the overall phase of the field.

As a result, the amplitude of the EM field at the observer plane after lens 2 at $z_{\tt r}\ll z_{12}$ from it, following the structure of (\ref{eq:amp00}), has the form similar to that of (\ref{eq:amp00}) and is given as below (see (\ref{eq:amp00=})--(\ref{eq:amp-Am=}) for the complete structure)
{}
\begin{eqnarray}
A_{\tt 2GL}(\vec x)=A_{02}(\vec x)F_{\tt 2GL}(\vec x),
  \label{eq:amp002}
\end{eqnarray}
where the wave amplitude at the observer, $A_{02}(\vec x)$, is given as  
{}
\begin{eqnarray}
A_{02}(\vec x)&=&
\frac{E_0}{z_{\tt t}+z_{12}+z_{\tt r}}
\exp\Big[i k\big(z_{\tt t}+z_{12}+z_{\tt r}\big)+i\hat\varphi_{1}\Big],
\label{eq:amp2}
\end{eqnarray}
where $\hat \varphi_{1}=kr_{g_1}\ln[2e(z_{\tt t}+z_{12})/r_{g_1}]-{\textstyle\frac{1}{4}}\pi$ collects the constant phase terms due to lens 1. With $A_{02}$ from (\ref{eq:amp2}), the amplification factor $F_{\tt 2GL}(\vec x)$ in (\ref{eq:amp002}) is given as follows 
{}
\begin{eqnarray}
F_{\tt 2GL}(\vec x) &=&
 \frac{z_{\tt t}+z_{12}+z_{\tt r}}{(z_{\tt t}+z_{12})z_{\tt r}}
 \frac{k}{2\pi i}\iint d^2\vec b_2 \,{\cal F}_{\tt 1GL}(\vec b_2)\exp\Big[ik\Big(\frac{b_2^2}{2  (z_{\tt t}+z_{12})}+\frac{1}{2 z_{\tt r}}({\vec b}_2 - \vec x)^2-2 r_{g_2}\ln(k b_2)\Big)\Big].
 \label{eq:amp-A2*10}
\end{eqnarray}
(Also see Appendix~\ref{sec:path-int} for a path integral derivation of (\ref{eq:amp-A2*10}).) Next, similarly to $\tilde z_1$ from (\ref{eq:amp-A1*}), we introduce another effective distance $\tilde z_2$ as below
{}
\begin{eqnarray}
 \frac{1}{\tilde z_2}= \frac{1}{z_{\tt t}+z_{12}}+\frac{1}{z_{\tt r}}\qquad \Rightarrow \qquad
 \tilde z_2=\frac{(z_{\tt t}+z_{12})z_{\tt r}}{z_{\tt t}+z_{12}+z_{\tt r}}.~~~
  \label{eq:r2tilde}
\end{eqnarray}
With $\tilde z_2$ from (\ref{eq:r2tilde}), we use expression for ${\cal F}_{\tt 1GL}(\vec b_2)$ from (\ref{eq:amp-A1*25}), remove the spherical wave (as in (\ref{eq:amp-A1*23})), and  present (\ref{eq:amp-A2*10}) as below 
{}
\begin{eqnarray}
F_{\tt 2GL}(\vec x) &=&
\sqrt{2\pi k r_{g_1}}  
 \frac{k}{i \tilde z_{2}}
\frac{1}{2\pi} \int_0^\infty \hskip -5pt
b_2db_2 \,J_0\big(k\theta_{1}b_2\big)\exp\Big[ik\Big(\frac{b_2^2}{2 \tilde z_{2}}-2 r_{g_2}\ln(k b_2)\Big)\Big]\int_0^{2\pi}\hskip -5pt d\phi_{\xi_2} 
 \exp\Big[-ik\frac{b_2\rho}{ z_{\tt r}}\cos[\phi_{\xi_2}-\phi]\Big]=\nonumber\\
 &=&
  \sqrt{2\pi k r_{g_1}}  
 \frac{k}{i \tilde z_{2}}
  \int_0^\infty b_2db_2 \,J_0\big(k\theta_{1}b_2\big)J_0\Big(k\frac{b_2\rho}{  z_{\tt r}}\Big)\exp\Big[ik\Big(\frac{b_2^2}{2 \tilde z_{2}}-2 r_{g_2}\ln(k b_2)\Big)\Big].~~~
  \label{eq:amp-A2*1}
\end{eqnarray}

Similarly to (\ref{eq:amp-A1*23}), we take  the integral over $b_2$ by the method of stationary phase, to determine $b_2=\sqrt{2r_{g_2}\tilde z_2}$ and, thus, the amplitude of the EM field on the image plane at the distance of $z_{\tt r}$ from the second lens is now given as  
{}
\begin{eqnarray}
F_{\tt 2GL}(\vec x)  &=&
\sqrt{2\pi k r_{g_1}}  \sqrt{2\pi k r_{g_2}}  e^{i\hat \varphi_{2}}
J_0\Big(k\frac{\sqrt{2r_{g_1} \tilde z_1}\sqrt{2r_{g_2} \tilde z_2}}{z_{12}}\Big)J_0\Big(k \frac{\sqrt{2r_{g_2} \tilde z_2}}{z_{\tt r}}\rho\Big),
  \label{eq:amp-A2*3=}
\end{eqnarray}
with  $\hat \varphi_{2}=kr_{g_2}\ln[2e(z_{\tt t}+z_{12}+z_{\tt r})/r_{g_2}] -{\textstyle\frac{1}{4}}\pi$, where we collected all the constant phase terms due to the presence of lens 2 on the eikonal, including the relevant $\phi_{\tt G2}$ term in (\ref{eq:amp-Am+}). (See Appendix~\ref{sec:alt-der} for alternative evaluation of (\ref{eq:amp-A2*1}).)  

Next, we use solution (\ref{eq:amp-A2*3=}), to determine the PSF of the two-thin-lens system 
{}
\begin{eqnarray}
{\tt PSF}_{\tt 2GL} (\vec x) &=&(2\pi k r_{g_1}) \, (2\pi k r_{g_2})  J^2_0\Big(k\frac{\sqrt{2r_{g_1} \tilde z_1}\sqrt{2r_{g_2} \tilde z_2}}{z_{12}}\Big)J^2_0\Big(k \frac{\sqrt{2r_{g_2} \tilde z_2}}{z_{\tt r}}\rho\Big),
  \label{eq:amp-PSF2=}
\end{eqnarray}
where we remind that in the two-lens case the distance to the receiver $z_{\tt r}$ from the one-lens case (\ref{eq:amp-A1*}) is replaced with $z_{12}$, so that the effective distance $ \tilde z_1$  has the form $ \tilde z_1={z_{\tt t}z_{12}}/({z_{\tt t}+z_{12}}).$

We observe that the argument of the first Bessel function in this expression is very large and is evaluated to be $k{\sqrt{2r_{g_1} \tilde z_1}\sqrt{2r_{g_2} \tilde z_2}}/{z_{12}} \geq 9.86\times 10^{6}\,(1~\mu{\rm m}/\lambda)(M_1/M_\odot)^\frac{1}{2}(M_2/M_\odot)^\frac{1}{2}(z_{\tt t}/650~{\rm AU})^\frac{1}{2}(z_{\tt r}/650~{\rm AU})^\frac{1}{2}(10~{\rm pc}/z_{12})$. In this case, the first Bessel function, $J^2_0(kb_1b_2/z_{12})$, can be approximated by using its expression for large arguments \cite{Abramovitz-Stegun:1965}:
 {}
\begin{eqnarray}
J^2_0\Big(k\frac{\sqrt{2r_{g_1} \tilde z_1}\sqrt{2r_{g_2} \tilde z_2}}{z_{12}}\Big)=\frac{1}{\sqrt{2\pi k r_{g_1}} \sqrt{2\pi k r_{g_2}} }\frac{z_{12}}{\sqrt{\tilde z_1 \tilde z_2}}\Big(1+\sin\big[\delta\varphi (\tilde z_1,\tilde z_2)\big]\Big), 
\quad 
{\rm with}
\quad 
\delta\varphi=2k\frac{\sqrt{2r_{g_1} \tilde z_1}\sqrt{2r_{g_2} \tilde z_2}}{z_{12}}.~
  \label{eq:bf0=}
\end{eqnarray}
Considering here the  term containing $\sin[\delta\varphi (\tilde z_1,\tilde z_2)]$, we note that at optical wavelengths, this term is a rapidly oscillating function of $\tilde z_{1}$ and $\tilde z_{2}$ that averages to 0. Therefore, the last term in the form of $J_0^2(x)$ expansion in (\ref{eq:bf0=}) may be neglected, allowing us to present the 
PSF from (\ref{eq:amp-PSF2=}) as below:
{}
\begin{eqnarray}
\left<{\tt PSF}_{\tt 2GL} (\vec x)\right> &=&\sqrt{2\pi k r_{g_1}}  \sqrt{2\pi k r_{g_2}}\frac{z_{12}}{\sqrt{\tilde z_1 \tilde z_2}}J^2_0\Big(k \frac{\sqrt{2r_{g_2} \tilde z_2}}{z_{\tt r}}\rho\Big).
  \label{eq:amp-psf2-av=+}
\end{eqnarray}

This is our main result for the PSF for a two-lens axially-arranged system of monopole gravitational lenses.
Note that result (\ref{eq:amp-psf2-av=+}) contains similar mass contributions from each of the lenses at both ends of the transmission link, $r_{g_1},r_{g_2}$. The amplification scales with the factor $z_{12}/\sqrt{\tilde z_1 \tilde z_2}$ involving the distance between the lenses, $z_{12}$, as well as the distances of both transmitter and receiver with respect to the transmitting and receiving lenses, $\tilde z_1, \tilde z_2$. 

This result yields the maximum amplification factor for the two-lens system with $z_{\tt t}, z_{\tt r}\ll z_{12}$:
{}
\begin{eqnarray}
\mu_{\tt 2GL} &=&\left<{\tt PSF}_{\tt 2GL} ( 0)\right>
\simeq \sqrt{2\pi k r_{g_1}}  \sqrt{2\pi k r_{g_2}}\frac{z_{12}}{\sqrt{ z_{\tt t}  z_{\tt r}}} 
\simeq \nonumber\\
&\simeq& 3.70\times 10^{14} \Big(\frac{M_1}{M_\odot}\Big)^\frac{1}{2}\Big(\frac{M_2}{M_\odot}\Big)^\frac{1}{2}\Big(\frac{1\,\mu{\rm m}}{\lambda}\Big)\Big(\frac{z_{12}}{10\,{\rm pc}}\Big)\Big(\frac{650\,{\rm AU}}{z_{\tt t}}\Big)^\frac{1}{2}
\Big(\frac{650\,{\rm AU}}{z_{\tt r}}\Big)^\frac{1}{2}.
  \label{eq:amp-A2*4=}
\end{eqnarray}
Clearly, for lensing architectures involving lenses with similar masses $r_{g_2} \simeq r_{g_1}$, the overall gain behaves as that of a single lens (\ref{eq:amp-PSF1}) scaled with the geometric factor of $z_{12}/\sqrt{ z_{\tt t}  z_{\tt r}}$, which provides additional gain for the lensing pair. 

Based on (\ref{eq:amp-psf2-av=+}), an observer in the focal region of  lens 2, positioned on the optical axis at the distance of $z_{\tt r}\geq R_2^2/2r_{g_2}$ from it, will see one Einstein ring around the lens with the radius $\theta_2$ given as:
{}
\begin{eqnarray}
\theta_2=\frac{\sqrt{2r_{g_2} \tilde z_2}}{z_{\tt r}}\simeq \sqrt{\frac{2r_{g_2}}{z_{\tt r}}}
 \simeq 7.80 \times 10^{-6}\,\Big(\frac{M_2}{M_\odot}\Big)^\frac{1}{2}\Big(\frac{650\,{\rm AU}}{z_{\tt r}}\Big)^\frac{1}{2}~ {\rm rad},
  \label{eq:theta1-r=}
\end{eqnarray}
which has the same structure as in the receiving lens case when a single lens is a part of a receiver (\ref{eq:theta1-r}). 

However, we note that the estimate (\ref{eq:amp-A2*4=}) may be misleading as it pertains only to the case when the receiver aperture is smaller than the first zero of the Bessel function $J_0\big(k \theta_2 \rho\big)$ present in (\ref{eq:amp-psf2-av=+}), namely $\rho_{\tt 2GL}=2.40483/(k \theta_2)\simeq 4.91\times 10^{-2}\,({\lambda}/{1\,\mu{\rm m}})({M_\odot}/{M_2})^\frac{1}{2}({z_{\tt r}}/{650\,{\rm AU}})^\frac{1}{2}\,{\rm m}$, as shown in Fig.~\ref{fig:psf-12}, which for optical wavelengths is not practical and will be averaged by the aperture, as in (\ref{eq:mu_av}). Therefore, for a realistic telescope, in context of a two-lens system, similarly to (\ref{eq:mu_av}), we develop an aperture-averaged value of the PSF, yielding the appropriate light amplification factor 
{}
\begin{eqnarray}
\left<\mu_{\tt 2GL}\right>&=&
\frac{1}{\pi ({\textstyle\frac{1}{2}}d_{\tt r})^2}
\int_0^{{\textstyle\frac{1}{2}}d_{\tt r}}\int_0^{2\pi}
{\tt PSF}_{\tt 2GL}({\vec x})\,\rho d\rho d\phi
\nonumber\\
&=&
\sqrt{2\pi k r_{g_1}}\sqrt{2\pi k r_{g_2}}\,
\frac{z_{12}}{\sqrt{\tilde z_1 \tilde z_2}}
\Big[
J^2_0\big(k \theta_2{\textstyle\frac{1}{2}}d_{\tt r}\big)
+
J^2_1\big(k \theta_2{\textstyle\frac{1}{2}}d_{\tt r}\big)
\Big]
\nonumber\\
&\simeq&
\frac{4 \sqrt{2r_{g_1} \tilde z_1}}{d_{\tt r}}\,
\frac{z_{12} z_{\tt r}}{\tilde z_1\tilde z_2}
\simeq
\sqrt{\frac{2r_{g_1}}{z_{\tt t}}}\,
\frac{4z_{12}}{d_{\tt r}}
\nonumber\\
&=&
\left.\left<\mu^{\tt r}_{\tt 1GL}\right>\right|_{z_{\tt r}=z_{\tt t}}
\frac{z_{12}}{z_{\tt t}}
=
9.62\times 10^{12}
\Big(\frac{M_1}{M_\odot}\Big)^\frac{1}{2}
\Big(\frac{650\, {\rm AU}}{z_{\tt t}}\Big)^\frac{1}{2}
\Big(\frac{1~{\rm m}}{d_{\tt r}}\Big)
\Big(\frac{z_{12}}{10\,{\rm pc}}\Big).
\label{eq:mu_av=}
\end{eqnarray}

Note that after averaging, the dependence on the mass of the second lens is absent in (\ref{eq:mu_av=}). The reason behind this is that the small scale of the diffraction pattern in (\ref{eq:amp-psf2-av=+}) necessitated the aperture averaging, which caused the second mass $r_{g_2}$ to drop out. (This is analogous to (\ref{eq:S_z*6z-mudd+})--(\ref{eq:mu_av}), where the same procedure also led to removing  a factor of $\sqrt{r_{g_1}}$.) This is because the telescope's size, $d_{\tt r}$, is much larger than the first zero of the diffraction pattern in (\ref{eq:amp-psf2-av=+}):
{}
\begin{eqnarray}
d_{\tt r}\gg \rho_{\tt 2GL}=2.40483/(k \theta_2)\simeq 0.38 \lambda \sqrt{\frac{z_{\tt r}}{2r_{g_2}}} \simeq 4.91\times 10^{-2}\Big(\frac{\lambda}{1\,\mu{\rm m}}\Big)\Big(\frac{M_\odot}{M_2}\Big)^\frac{1}{2}\Big(\frac{z_{\tt r}}{650\,{\rm AU}}\Big)^\frac{1}{2}~ {\rm m}.
  \label{eq:theta1-diff}
\end{eqnarray}

Note that when condition (\ref{eq:theta1-diff}) is not satisfied and the aperture-averaging may not be relevant (i.e., due to a larger wavelength, larger receiver distance, etc.), one would have to use the entire PSF with the $J_0^2$(x) factors included, as in the case (\ref{eq:S_z*6z-mu2+}). In these cases, the relevant PSFs are (\ref{eq:S_z*6z-mu2+}), (\ref{eq:S_z*6z-mudd+}) for one lens, and either (\ref{eq:amp-psf2-av=+}) or (\ref{eq:amp-psf2-av}) for two lenses. 

Comparing result (\ref{eq:mu_av=}) with the transmission scenarios involving a single lens (\ref{eq:amp-PSF1}) and (\ref{eq:mu_av}), we see that the two-lens system provides significant additional light amplification captured by the factor $(z_{12}/z_{\tt t})$. It is 82.5 times more effective compared to the transmitting lens case (\ref{eq:S_z*6z-mu2+})  and is $3.17\times 10^3$ times more effective than the receiving lens scenario (\ref{eq:mu_av}). Clearly, transmission via a pair of lenses benefits from gravitational amplification at both ends of the transmission link, thus enabling interstellar power transmission with modern-day optical instrumentation.

\section{Power transmission}
\label{sec:power-X}
 
To assess the effectiveness of power transmission using gravitational lensing, we consider three transmission scenarios that involve lensing with either a single lens or double lenses. These scenarios differ not only in the number of lenses used but also in the positions of the transmitter and receiver relative to the lenses.

\subsection{Single lens: transmission from its focal region}
\label{sec:single-A}

First, we consider transmission via a single lens with the transmitter positioned in its focal region at a distance of $z_{\tt t}\geq R^2_1/2r_{g_1}$ from the lens, while the receiver is at interstellar distance of $z_{\tt r}\gg z_{\tt t}$. This transmitting-lens scenario is case iii) in Sec.~\ref{sec:transmit-geom} and was addressed by (\ref{eq:S_z*6z-mu2+}). We consider the transmitter to be a point source. 

We assume that transmission is characterized by the beam divergence set by the telescope's aperture $d_{\tt t}$ yielding angular resolution of $\theta_0\simeq\lambda/d_{\tt t}=1.00 \times10^{-6}~(\lambda/1\,\mu{\rm m})(1\,{\rm m}/d_{\tt t})\,{\rm rad}$.
When the signal reaches the receiver at the distance of $(z_{\tt t}+z_{\tt r})$ from the transmitter, the beam is expanded to a large spot with the radius of $\rho_*=(z_{\tt t}+z_{\tt r})(\lambda/d_{\tt t})$. In addition, while passing by the lens, the light is amplified according to (\ref{eq:S_z*6z-mu2+}). As a result, a telescope with the aperture $d_{\tt r}$ receives a fraction of the transmit power that  is evaluated to be:
 {}
\begin{eqnarray}
P^{\tt t}_{\tt1GL}&=&
P_0 \, {\tt PSF}^{\tt t}_{\tt 1GL}({\vec x})
\frac{ \pi ({\textstyle\frac{1}{2}}d_{\tt r})^2}{\pi \rho_*^2}
\nonumber\\
&=&
P_0
\frac{ \pi({\textstyle\frac{1}{2}}d_{\tt r})^2}
{\pi(z_{\tt t}+z_{\tt r})^2}
\Big(\frac{d_{\tt t}}{\lambda}\Big)^2
\,2\pi k r_{g_1}\,
J^2_0\Big(k\frac{\sqrt{2r_{g_1} z_{\tt t}}}{ z_{\tt r}}\rho \Big)
\nonumber\\
&\simeq&
3.06 \times 10^{-13}
\Big(\frac{P_0}{1~{\rm W}}\Big)
\Big(\frac{1~\mu{\rm m}}{\lambda}\Big)^3
\Big(\frac{d_{\tt t}}{1~{\rm m}}\Big)^2
\Big(\frac{d_{\tt r}}{1~{\rm m}}\Big)^2
\Big(\frac{10~{\rm pc}}{z_{\tt r}}\Big)^2
\Big(\frac{M_1}{M_\odot}\Big)~~~ {\rm W}.
\label{eq:powed-l}
\end{eqnarray}   

For the same transmitting scenario, a free space laser power transmission in the vacuum is described as  
 {}
\begin{eqnarray}
P_{\tt free}&=& \frac{P_0 \pi ({\textstyle\frac{1}{2}}d_{\tt r})^2}{\pi (z_{\tt t}+z_{\tt r})^2} \Big(\frac{d_{\tt t}}{\lambda}\Big)^2 
\simeq 
2.62 \times 10^{-24} \Big(\frac{P_0}{1~{\rm W}}\Big)\Big(\frac{1~\mu{\rm m}}{\lambda}\Big)^2
\Big(\frac{d_{\tt t}}{1~{\rm m}}\Big)^2
\Big(\frac{d_{\tt r}}{1~{\rm m}}\Big)^2\Big(\frac{10~{\rm pc}}{z_{\tt r}}\Big)^2
~~ {\rm W}.~~~
\label{eq:po-free}
\end{eqnarray}

Comparing  results (\ref{eq:powed-l})  and (\ref{eq:po-free}), we observe that signal transmission relying on a single gravitational lens amplifies the received power by $P^{\tt t}_{\tt1GL}/P_{\tt free} \simeq 1.17\times 10^{11}$, as prescribed by (\ref{eq:amp-PSF1}). 
We note that, depending on the transmitter's performance, the power amplification value (\ref{eq:powed-l})  would have to be adjusted to account for a realistic system's throughput. Below, we consider several   transmitter implementation approaches relevant to this scenario.  

First, from a practical standpoint, we observe that a single transmitter with the combination of parameters $d_{\tt t}$ and $\lambda$ chosen in (\ref{eq:powed-l}), will be able to form a beam only with very short impact parameters of $\rho_{\tt mult}\simeq z_{\tt t}{\textstyle\frac{1}{2}}\theta_0 =z_{\tt t}(\lambda/2 d_{\tt t}) =0.07 R_\odot \, (\lambda/1\,\mu{\rm m})(1\,{\rm m}/d_{\tt t})(z_{\tt t}/650\,{\rm AU})(R_1/R_\odot)$. The corresponding EM field will be totally absorbed by the lens.  

Therefore, the parameter choice (\ref{eq:powed-l}) would require a special transmitter design that would rely on multiple laser transmitter heads. Considering the Einstein ring that is formed with the radius of $R_{\tt ER}=\sqrt{2r_{g_1} z_{\tt t}}$, one choice would be to take $n_{\tt t}=2\pi R_{\tt ER}/2 \rho_{\tt mult}= 2\pi (d_{\tt t}/\lambda)\sqrt{2r_{g_1}/z_{\tt t}}\simeq 48.98\, (d_{\tt t}/1\,{\rm m})(1\,\mu{\rm m}/\lambda)\,(M_1/M_\odot)^\frac{1}{2}(650\,{\rm AU}/z_{\tt t})^\frac{1}{2}$ laser heads arranged in a conical shape each with the same offset angle of  $R_{\tt ER}/z_{\tt t}= \sqrt{2r_{g_1}/z_{\tt t}}\simeq 1.61'' \,(M_1/M_\odot)^\frac{1}{2}(650\,{\rm AU}/z_{\tt t})^\frac{1}{2}$ from the mean direction to the lens. In effect, such a transmitter will illuminate the Einstein ring on the circumference of the lens with the signal of total power $n_{\tt t} P_0$ that will then proceed toward the receiver.  

In the scenario with multiple transmitting heads, the role of the transmitter architecture is to synthesize the annular illumination assumed in the idealized link budget (\ref{eq:powed-l}). A fully self-consistent finite-aperture calculation for a discrete set of off-axis transmitting apertures is beyond the scalar point-source model used here. We therefore parameterize the architecture-dependent optical throughput by $\eta^{\tt t}_{\tt ann}\leq 1$. In the ideal annular-illumination limit, the received power is
{}
\begin{eqnarray}
P^{\tt t}_{{\tt1GL},{\tt ann}}
&=&
\eta^{\tt t}_{\tt ann}\,
P_0\,{\tt PSF}^{\tt t}_{\tt 1GL}(\vec x)
\frac{\pi({\textstyle\frac{1}{2}}d_{\tt r})^2}
{\pi(z_{\tt r}\lambda/d_{\tt t})^2}
\nonumber\\
&=&
\eta^{\tt t}_{\tt ann}\,
P_0\,2\pi k r_{g_1}
\Big(\frac{d_{\tt t}}{\lambda}\Big)^2
\frac{d_{\tt r}^2}{4z_{\tt r}^2}
J^2_0\Big(k\frac{\sqrt{2r_{g_1} z_{\tt t}}}{z_{\tt r}}\rho\Big).
\label{eq:powed-l-mult}
\end{eqnarray}
For $\eta^{\tt t}_{\tt ann}=1$, Eq.~(\ref{eq:powed-l-mult}) is identical to (\ref{eq:powed-l}) in the on-axis, $z_{\tt t}\ll z_{\tt r}$ limit. 

Next, a single transmitter with a small aperture, for example $d^{\tt sm}_{\tt t}=6$~cm, may illuminate impact parameters comparable to the stellar radius:
\[
p_{\tt sing}=z_{\tt t}\frac{\lambda}{2d^{\tt sm}_{\tt t}}
\simeq
1.16 R_\odot
\Big(\frac{\lambda}{1\,\mu{\rm m}}\Big)
\Big(\frac{6\,{\rm cm}}{d^{\tt sm}_{\tt t}}\Big)
\Big(\frac{z_{\tt t}}{650\,{\rm AU}}\Big).
\]
For a hard solar occulting disk, the geometrically blocked fraction would be approximately $(R_\odot/p_{\tt sing})^2\simeq 0.74$ for these fiducial numbers, leaving about $0.26$ of the beam power outside the photospheric disk before additional losses due to the corona, beam apodization, and the desired annular illumination pattern are included.  Because those losses depend on the specific transmitter design, we do not assign a universal throughput factor to this single-aperture option.

Finally, going back to the multi-head transmitter design discussed above, we note that, using a transmitter with $n_{\tt t}\sim49$ laser heads, each pointing in a different direction, may be complicated and other designs should be considered. As an example, one may consider a holographic diffusing element\footnote{As shown here:
\url{https://www.rpcphotonics.com/engineered-diffusers-information}.} that can be used to sculpt an outgoing beam to practically any shape. Such a diffuser may allow for a uniform illumination of the Einstein ring at a specified angular separation from the lens, thus offering a plausible approach to a transmitter design. In this case, there will be no significant loss in the optical throughput and the entire power amplification of (\ref{eq:powed-l}) will be at work.  

We observe that each of the three transmission architectures above are different, yielding different optical transmission throughputs, specific noise contributions, and will be characterized by different SNR performances \cite{Turyshev-Toth:2022a,Turyshev-Toth:2022b}.  In any case, these types of the transmission scenarios  may be used to search for the signals before moving to the focal region of lens 2 (discussed in Sec.~\ref{sec:double-C}) that would be needed to establish a reliable communication infrastructure. 

\subsection{Single lens: reception at its focal region}
\label{sec:single-B}

Another single-lens transmission scenario involves transmission from an interstellar distance into the focal region of a lens, corresponding to case iv) of Sec.~\ref{sec:transmit-geom}.  The transmitter is at a large distance $z_{\tt t}$ from the lens, while the receiver is in the focal region at $z_{\tt r}\geq R_1^2/(2r_{g_1})$, with $z_{\tt t}\gg z_{\tt r}$.  This case is covered by Eq.~(\ref{eq:mu_av}).  The only difference from the scenario discussed in Sec.~\ref{sec:single-A} is that the laser beam divergence in this case will naturally result in a large spot size illuminated by the transmitter at the lens plane, thus reducing the incident power and overall link performance. 

We note that a telescope with aperture $d_{\tt r}$ positioned on the optical axis in the focal area of a lens and looking back at it would see the Einstein ring formed around the lens. The energy deposited in the ring will be the same as that received by the observer. This means that the effective collecting area of the receiving telescope $A_{\tt tel}=\pi ({\textstyle\frac{1}{2}}d_{\tt r})^2 \left<\mu_{\tt 1GL}^{\tt r}\right>$, with $\left<\mu_{\tt 1GL}^{\tt r}\right>$ from (\ref{eq:mu_av}), is equal to the  area subtended by the observed Einstein ring  $A_{\tt ER}=2\pi R_{\tt ER}w_{\tt ER}$, where $R_{\tt ER}=\sqrt{2r_{g_1} z_{\tt r}}$. Equating $A_{\tt tel}=A_{\tt ER}$, allows us to determine the width of the ring $w_{\tt ER} ={\textstyle\frac{1}{2}}d_{\tt r}$, thus establishing the effective area subtended by the Einstein ring as seen by the receiving telescope in the case of a single lens transmission
{}
\begin{eqnarray}
A_{\tt ER1}=\pi d_{\tt r} \sqrt{2r_{g_1} z_{\tt r}}.
  \label{eq:area-ER}
\end{eqnarray}

Clearly, a telescope with the angular resolution of $1.22(\lambda/d_{\tt r})=1.22\times 10^{-6} (\lambda/1\,\mu{\rm m})(1\,{\rm m}/d_{\tt r})\gg d_{\tt r}/2z_{\tt r}=5.14\times 10^{-15}(d_{\tt r}/1\,{\rm m})(650\,{\rm AU}/z_{\tt r})$, will not be able to resolve the width of the ring; however, the length of its circumference is resolved with $n_{\tt t}=2\pi \sqrt{2r_{g_1}/ z_{\tt r}}/(\lambda/d_{\tt r})\simeq 48.98 \,({M_1}/{M_\odot})^\frac{1}{2}({650~{\rm AU}}/{z_{\tt r}})^\frac{1}{2}(1~\mu{\rm m}/\lambda)({d_{\tt r}}/{1~{\rm m}})$ resolution elements. This information is useful when considering transmitter design or evaluating receiver performance.  

To evaluate the effectiveness of such a scenario, consider a transmitting telescope that illuminates the Einstein ring around the lens at the radius of $R_{\tt ER}=\sqrt{2r_{g_1} z_{\tt r}}$. However, because of diffraction, the transmitted energy will spread over a much larger area with the radius of  $\rho_*'=z_{\tt t}(\lambda/d_{\tt t})=443.54 R_\odot \, (\lambda/1\,\mu{\rm m})(z_{\tt t}/10\,{\rm pc})(1\,{\rm m}/d_{\tt t})(R_1/R_\odot)$.  Thus, not all the energy will be received at the Einstein ring, so some of the energy will be lost.  By the time the signal reaches the receiver, its deviation from the optical axis is controlled by gravity and changes by $z_{\tt r}\theta^{\tt r}_1$, where $\theta^{\tt r}_1$ is from (\ref{eq:theta1-r}). As a result, we estimate the total received power in this case when the single lens acts as part of a receiver as below
 {}
\begin{eqnarray}
P^{\tt r}_{\tt1GL}&=&P_0  \frac{A_{\tt ER1}}{\pi (z_{\tt t}(\lambda/d_{\tt t}))^2}  \frac{\pi R_{\tt ER}^2}{\pi (z_{\tt r}\theta^{\tt r}_1)^2} =
\frac{P_0 \pi ({\textstyle\frac{1}{2}}d_{\tt r})^2}{\pi z_{\tt t}^2}  \Big(\frac{d_{\tt t}}{\lambda}\Big)^2 \frac{4 \sqrt{2r_{g_1} z_{\tt r}}}{d_{\tt r}}=\nonumber\\
&=&
7.96 \times 10^{-15}\Big(\frac{P_0}{1~{\rm W}}\Big)\Big(\frac{1~\mu{\rm m}}{\lambda}\Big)^2
\Big(\frac{d_{\tt t}}{1~{\rm m}}\Big)^2
\Big(\frac{d_{\tt r}}{1~{\rm m}}\Big)\Big(\frac{10~{\rm pc}}{z_{\tt t}}\Big)^2\Big(\frac{M_1}{M_\odot}\Big)^\frac{1}{2}\Big(\frac{z_{\tt r}}{650\,{\rm AU}}\Big)^\frac{1}{2}~~~ {\rm W},
\label{eq:powed-l02}
\end{eqnarray}
which is a factor 38.47 times smaller than that obtained in the scenario with transmission from the focal region (\ref{eq:powed-l}) discussed in Sec.~\ref{sec:single-A}.  As a result, there is a difference between transmitting from the focal region of a lens to a receiver at a large interstellar distance and transmitting from a large distance into the focal region of the lens.  In addition, there may be different noise sources involved affecting the SNR performance of a transmission link. 

\subsection{Two lenses: transmission and reception from/to the focal regions}
\label{sec:double-C}

Now we consider a two-lens power-transmission scenario.  The transmitter is positioned in the focal region of lens 1 at a distance $z_{\tt t}>R_1^2/(2r_{g_1})$ from it and sends a signal toward lens 2, which is separated from lens 1 by the interstellar distance $z_{12}\gg z_{\tt t}$.  After passing lens 2, the signal is detected by a receiver positioned in the focal region of lens 2 at a distance $z_{\tt r}>R_2^2/(2r_{g_2})$ from it. Analytically, this case is described by solution (\ref{eq:amp-mu2}). 

The power transmission, as outlined above, is characterized by (\ref{eq:powed-l}), where the light amplification is from (\ref{eq:amp-mu2}). In this case, we follow the same approach that we used to derive (\ref{eq:powed-l}). We may treat the two-lens system in the same manner as we dealt with a single-lens transmission (\ref{eq:powed-l}).  For that, we realize that when the signal reaches the receiver at the distance of $(z_{\tt t}+z_{12}+z_{\tt r})$, the beam is expanded to a large spot with the radius of $\rho_{**}=(z_{\tt t}+z_{12}+z_{\tt r})(\lambda/d_{\tt t})\simeq z_{12}(\lambda/d_{\tt t})$. In addition, while passing by the two-lens system, the light is amplified according to (\ref{eq:amp-mu2}).  

As a result, a telescope with the aperture $d_{\tt r}$ receives a fraction of the transmit power that  is evaluated to be:
 {}
\begin{eqnarray}
P_{\tt2GL}&=&
P_0 \left<\mu_{\tt 2GL}\right>
\frac{ \pi ({\textstyle\frac{1}{2}}d_{\tt r})^2}{\pi \rho_{**}^2}
=
\frac{P_0 \pi ({\textstyle\frac{1}{2}}d_{\tt r})^2}{\pi z_{12}^2}
\Big(\frac{d_{\tt t}}{\lambda}\Big)^2
\sqrt{\frac{2r_{g_1}}{z_{\tt t}}}\frac{4z_{12}}{d_{\tt r}}
\nonumber\\
&\simeq&
2.53 \times 10^{-11}
\Big(\frac{P_0}{1~{\rm W}}\Big)
\Big(\frac{1~\mu{\rm m}}{\lambda}\Big)^2
\Big(\frac{d_{\tt t}}{1~{\rm m}}\Big)^2
\Big(\frac{d_{\tt r}}{1~{\rm m}}\Big)
\Big(\frac{10~{\rm pc}}{z_{12}}\Big)
\Big(\frac{M_1}{M_\odot}\Big)^\frac{1}{2}
\Big(\frac{650\,{\rm AU}}{z_{\tt t}}\Big)^\frac{1}{2}~~~ {\rm W}.
\label{eq:powed-l2}
\end{eqnarray}
which clearly demonstrates the benefits of the gravitational amplification provided by the double-lens systems while yielding the gain of $P_{\tt2GL}/P_{\tt free}=\sqrt{{2r_{g_1}}/{z_{\tt t}}}(4z_{12}/d_{\tt r})\simeq 9.62\times 10^{12}\,({M_1}/{M_\odot})^\frac{1}{2}({650\, {\rm AU}}/{z_{\tt t}})^\frac{1}{2}({1~{\rm m}}/{d_{\tt r}})({z_{12}}/{10~{\rm pc}})$, which is $\sim (2/\pi^2) (\lambda/d_{\tt r})(z_{12}/\sqrt{2r_{g_1}z_{\tt t}})\simeq 82.50 \,({\lambda}/{1~\mu{\rm m}})({1~{\rm m}}/{d_{\tt r}})({M_\odot}/{M_1})^\frac{1}{2}({650\, {\rm AU}}/{z_{\tt t}})^\frac{1}{2}({z_{12}}/{10~{\rm pc}})$ times higher than in the case when a single lens acts as a part of a transmitter, shown by  (\ref{eq:powed-l}), or $z_{12}/z_{\tt t}\simeq 3.17\times 10^3\,({z_{12}}/{10~{\rm pc}})({650\, {\rm AU}}/{z_{\tt t}})$ times higher than that of a single lens acting as a part of the receiver as seen by (\ref{eq:powed-l02}).

We may consider an alternative derivation for the two-lens transmission link (\ref{eq:powed-l2}). A telescope with aperture $d_{\tt r}$ would see two unresolved Einstein-ring contributions around the lens.  The energy deposited in the ring is the same as that received by the telescope. This means that the effective collecting area of the receiving telescope, $A_{\tt tel}=\pi ({\textstyle\frac{1}{2}}d_{\tt r})^2 \left<\mu_{\tt 2GL}\right>$, with $\left<\mu_{\tt 2GL}\right>$ from (\ref{eq:amp-mu2}), is equal to the area subtended by the observed Einstein ring, $A_{\tt ER}=2\pi R_{\tt ER}w_{\tt ER}$, where $R_{\tt ER}=\sqrt{2r_{g_1} z_{\tt t}}$. Equating $A_{\tt tel}=A_{\tt ER}$ allows us to determine the width of the ring, $w_{\tt ER} ={\textstyle\frac{1}{2}}d_{\tt r}(z_{12}/z_{\tt t})$, thus establishing the effective area subtended by the Einstein ring as seen by the receiving telescope in the case of a two-lens transmission
{}
\begin{eqnarray}
A_{\tt ER2}
&=&
2\pi\sqrt{2r_{g_1}z_{\tt t}}\,
\Big[{\textstyle\frac{1}{2}}d_{\tt r}\frac{z_{12}}{z_{\tt t}}\Big]
=
\pi d_{\tt r} z_{12}\sqrt{\frac{2r_{g_1}}{z_{\tt t}}}
\nonumber\\
&=&
\left.A_{\tt ER1}\right|_{z_{\tt r}=z_{\tt t}}\frac{z_{12}}{z_{\tt t}}.
\label{eq:area-ER2}
\end{eqnarray}
where $A_{\tt ER1}$ from (\ref{eq:area-ER}) is evaluated at $z_{\tt r}=z_{\tt t}$. Therefore, the amount of light captured by the ring increases with distance as $(z_{12}/z_{\tt t})$, as does the amplification factor (\ref{eq:amp-mu2}). This factor is the key source of additional amplification for two-lens transmissions. 

For the fiducial geometry, the angular width of this effective ring is
\[
\frac{w_{\tt ER}}{z_{\tt r}}
=
\frac{d_{\tt r}z_{12}}{2z_{\tt t}z_{\tt r}}
\simeq
1.6\times 10^{-11}
\Big(\frac{d_{\tt r}}{1\,{\rm m}}\Big)
\Big(\frac{z_{12}}{10\,{\rm pc}}\Big)
\Big(\frac{650\,{\rm AU}}{z_{\tt t}}\Big)
\Big(\frac{650\,{\rm AU}}{z_{\tt r}}\Big),
\]
well below the diffraction resolution $1.22\lambda/d_{\tt r}$ of a metre-class optical telescope. 

The same annular-illumination interpretation applies to the two-lens configuration. A detailed finite-aperture calculation for a discrete transmitter array is not developed here; instead, we assume that the transmitter architecture synthesizes the annular field required by the two-lens PSF and write any architecture-dependent throughput as $\eta_{\tt 2GL}\leq1$. In the ideal limit, the beam occupies a diffraction-limited patch of characteristic radius $z_{12}\lambda/d_{\tt t}$ over the interstellar separation, giving
{}
\begin{eqnarray}
P_{{\tt2GL},{\tt ann}}
&=&
\eta_{\tt 2GL}\,
P_0\left<\mu_{\tt 2GL}\right>
\frac{\pi({\textstyle\frac{1}{2}}d_{\tt r})^2}
{\pi(z_{12}\lambda/d_{\tt t})^2}
\nonumber\\
&=&
\eta_{\tt 2GL}\,
\frac{P_0\pi({\textstyle\frac{1}{2}}d_{\tt r})^2}{\pi z_{12}^2}
\Big(\frac{d_{\tt t}}{\lambda}\Big)^2
\sqrt{\frac{2r_{g_1}}{z_{\tt t}}}\frac{4z_{12}}{d_{\tt r}} .
\label{eq:powed-2GL}
\end{eqnarray}
For $\eta_{\tt 2GL}=1$, Eq.~(\ref{eq:powed-2GL}) is identical to (\ref{eq:powed-l2}). 
As a result, in the case of the two-lens transmission, the total power received is higher than that for a single-lens transmission.  In other words, the second lens adds more power by focusing light, as expected.

\section{Detection sensitivity} 
\label{sec:SNR}

Now that we have established the power estimates for the transmission links characteristic for various lensing architectures involving one and two lenses, we need to consider practical aspects of such transmission links. For this, we evaluate the contribution of the brightness of the stellar atmospheres to the overall detection sensitivity. The signal estimates are summarized in Sec.~\ref{sec:signal-estimates}, and the background models are discussed in Sec.~\ref{sec:sol-corona-model}.  

\subsection{Relevant signal estimates}
\label{sec:signal-estimates}

First of all, based on the results for power transmission for various configurations considered, namely (\ref{eq:powed-l}), for a lens near the transmitter,  (\ref{eq:powed-l02}), for a lens on a receiving end,  and (\ref{eq:powed-l2}) for the pair of lenses, we have the following estimates for photon flux, $Q^{\tt t}_{\tt1GL}, Q^{\tt r}_{\tt1GL}, Q_{\tt2GL}$, received by the telescope in all three of these cases
{}
\begin{eqnarray}
Q^{\tt t}_{\tt1GL}&=&\frac{\lambda}{hc}P^{\tt t}_{\tt1GL}
\simeq
1.54 \times 10^{6}
\Big(\frac{P_0}{1~{\rm W}}\Big)
\Big(\frac{1~\mu{\rm m}}{\lambda}\Big)^2
\Big(\frac{d_{\tt t}}{1~{\rm m}}\Big)^2
\Big(\frac{d_{\tt r}}{1~{\rm m}}\Big)^2
\Big(\frac{10~{\rm pc}}{z_{\tt r}}\Big)^2
\Big(\frac{M_1}{M_\odot}\Big)~~~ {\rm phot/s},
\label{eq:powed-l-Q}
\\
Q^{\tt r}_{\tt1GL}&=&\frac{\lambda}{hc}P^{\tt r}_{\tt1GL}
\simeq
4.01 \times 10^{4}
\Big(\frac{P_0}{1~{\rm W}}\Big)
\Big(\frac{1~\mu{\rm m}}{\lambda}\Big)
\Big(\frac{d_{\tt t}}{1~{\rm m}}\Big)^2
\Big(\frac{d_{\tt r}}{1~{\rm m}}\Big)
\Big(\frac{10~{\rm pc}}{z_{\tt t}}\Big)^2
\Big(\frac{M_1}{M_\odot}\Big)^\frac{1}{2}
\Big(\frac{z_{\tt r}}{650\,{\rm AU}}\Big)^\frac{1}{2}~~~ {\rm phot/s},
\label{eq:powed-l02-Q}
\\
Q_{\tt2GL}&=&\frac{\lambda}{hc}P_{\tt2GL}
\simeq
1.27 \times 10^{8}
\Big(\frac{P_0}{1~{\rm W}}\Big)
\Big(\frac{1~\mu{\rm m}}{\lambda}\Big)
\Big(\frac{d_{\tt t}}{1~{\rm m}}\Big)^2
\Big(\frac{d_{\tt r}}{1~{\rm m}}\Big)
\Big(\frac{10~{\rm pc}}{z_{12}}\Big)
\Big(\frac{M_1}{M_\odot}\Big)^\frac{1}{2}
\Big(\frac{650\,{\rm AU}}{z_{\tt t}}\Big)^\frac{1}{2}~~~ {\rm phot/s}.
\label{eq:powed-l2-Q}
\end{eqnarray}

Therefore, based on the estimates in Sec.~\ref{sec:signal-estimates}, we have significant photon fluxes in each of the three configurations considered. However, each of the three configurations will have different contributions from the brightness of the stellar atmospheres along the relevant transmissions links. Below, we will provide estimates for the relevant noise. 

\subsection{Dominant noise sources}
\label{sec:sol-corona-model}

The Einstein rings that form around the stellar lenses appear against bright stellar photospheric and coronal backgrounds.  To assess these noise contributions, we estimate the photon rates from the lenses themselves and from their coronas in the annular regions where the Einstein rings appear.  Not all of this light can be blocked by coronagraphs, and its shot noise must be included in the sensitivity analysis.  Other backgrounds at small angular separations may also be important for specific systems, but they are not modeled in detail here. 

Below, we will focus on the main anticipated sources of noise -- those from lenses themselves and from their coronas. The three noise cases are treated in Secs.~\ref{sec:nosie-Xmit}--\ref{sec:nosie-2-lens}.

\subsubsection{Single lens: transmission from its focal region}
\label{sec:nosie-Xmit}

In the case, when the lens is at the transmitter node, the situation is quite different from the one that we considered in Sec.~\ref{sec:nosie-receiver}. In this case the angular extent of the Einstein ring is given by (\ref{eq:theta1-t}). The angular size of the stellar lens as seen by the distant receiver is $R_1/z_{\tt r}=2.26\times 10^{-9}(R_1/R_\odot)(10\,{\rm pc}/z_{\tt r})\,{\rm rad}$, similar to that of the Einstein ring. A conventional telescope with a practical diameter will not be able to resolve this star, therefore, the use of a coronagraph is out of the question. Thus, the light emitted by that stellar lens and received by the telescope will be the noise that must be dealt with. In other words, the light from the Einstein ring will arrive together with the unattenuated light from the star.  

We consider the Sun as a stand-in for lens 1. 
In this case, assuming the Sun's temperature\footnote{See \url{https://en.wikipedia.org/wiki/Sun}} to be $T_\odot=5\,772$~K, we estimate the solar brightness from Planck's radiation law:
{}
\begin{equation}
B_\odot=
\frac{\sigma T_\odot^4}{\pi}
=
\int_0^\infty B_\lambda(T_\odot)d\lambda
=
\int_0^\infty
\frac{2hc^2}{\lambda^5\big(e^{hc/\lambda k_B T_\odot}-1\big)}d\lambda
=
2.0034 \times 10^7 ~~ \frac{\rm W}{{\rm m}^2\,{\rm sr}},
\label{eq:model-L02*}
\end{equation}
where we use a blackbody radiation model with $\sigma$ the Stefan--Boltzmann constant and $k_B$ the Boltzmann constant. 
The bolometric radiance in (\ref{eq:model-L02*}) yields the total power output of the Sun of $L_\odot= \pi B_\odot 4\pi R^2_\odot=3.828 \times 10^{26}$ W, which is one nominal solar luminosity as defined by the IAU.\footnote{See \url{https://www.iau.org/static/resolutions/IAU2015_English.pdf}}

When dealing with laser light propagating in the vicinity of the Sun, we need to be concerned with the flux within some bandwidth $\Delta \lambda$ around the laser wavelength $\lambda$, assuming we can filter the light that falls outside $\Delta \lambda$, then 
{}
\begin{align}
B_\odot (\lambda, \Delta \lambda)
&{}=
B_\lambda(T_\odot)\Delta \lambda
= \frac{2hc^2}{\lambda^5\big(e^{hc/\lambda k_B T_\odot}-1\big)}\Delta \lambda
\simeq1.07 \times 10^5 \Big(\frac{1\,\mu{\rm m}}{\lambda}\Big)^5\Big(\frac{\Delta \lambda}{10\,{\rm nm}}\Big)~~   \frac{\rm W}{{\rm m}^2\,{\rm sr}}.
\label{eq:model-L0*}
\end{align}
We take (\ref{eq:model-L0*}) to derive the luminosity of the Sun within a narrow bandwidth as follows
{}
\begin{align}
L_\odot (\lambda, \Delta \lambda)
&{}=
4\pi^2 R_\odot^2 B_\odot(\lambda,\Delta\lambda)
=
\frac{8\pi^2 h c^2 R_\odot^2}{\lambda^5
\big(e^{hc/\lambda k_B T_\odot}-1\big)}\Delta \lambda
\nonumber\\
&\simeq
2.05 \times 10^{24}
\Big(\frac{1\,\mu{\rm m}}{\lambda}\Big)^5
\Big(\frac{\Delta \lambda}{10\,{\rm nm}}\Big)~~{\rm W}.
\label{eq:model-L0*t}
\end{align}

As a result, in the case when a lens is used as a part of the transmitter, the power received from the lensing star by a telescope and the corresponding photon flux are derived from (\ref{eq:model-L0*t})  and are given as 
{}
\begin{eqnarray}
P^{\tt t\, \star}_{\tt1GL}
&=&
L_1 (\lambda, \Delta \lambda)
\frac{\pi ({\textstyle\frac{1}{2}}d_{\tt r})^2}{4\pi z_{\tt r}^2}
\nonumber\\
&=&
1.35 \times 10^{-12}
\Big(\frac{L_1}{L_\odot}\Big)
\Big(\frac{1\,\mu{\rm m}}{\lambda}\Big)^5
\Big(\frac{\Delta \lambda}{10\,{\rm nm}}\Big)
\Big(\frac{d_{\tt r}}{1~{\rm m}}\Big)^2
\Big(\frac{10~{\rm pc}}{z_{\tt r}}\Big)^2~~~{\rm W},
\label{eq:model-PP-t}\\
Q^{\tt t\, \star}_{\tt1GL}
&=&
\frac{\lambda}{hc}P^{\tt t\, \star}_{\tt1GL}
\nonumber\\
&=&
6.78\times 10^6
\Big(\frac{L_1}{L_\odot}\Big)
\Big(\frac{1\,\mu{\rm m}}{\lambda}\Big)^4
\Big(\frac{\Delta \lambda}{10\,{\rm nm}}\Big)
\Big(\frac{d_{\tt r}}{1~{\rm m}}\Big)^2
\Big(\frac{10~{\rm pc}}{z_{\tt r}}\Big)^2~~~{\rm phot/s}.
\label{eq:pow-fp==2}
\end{eqnarray}
Equations~(\ref{eq:model-PP-t})--(\ref{eq:pow-fp==2}) give the unresolved stellar background for the transmitting-lens case.

\subsubsection{Single lens: reception at its focal region}
\label{sec:nosie-receiver}

The case of a single lens positioned nearby receiver is the most straightforward one. In this scenario, the angular size of the Einstein ring is given by (\ref{eq:theta1-r}), which is quite large. The receiving telescope with the diameter of $d_{\tt r}$ resolves the angular radius of the stellar lens in $(R_1/z_{\tt r})/1.22(\lambda/d_{\tt r})=5.86\,(R_1/R_\odot)(650\,{\rm AU}/z_{\tt r})(1\,\mu{\rm m}/\lambda)(d_{\tt r}/1\,{\rm m})$ resolution elements. Thus, one may consider using a coronagraph to block the stellar light (in our case, this would be the Sun.)  

Most of the relevant modeling was already accomplished in the context of our SGL studies. For that we know that the transmitted signal is seen by the telescope within the annulus that corresponds to the Einstein ring around a lens at the distance of $b=\sqrt{2r_{g_1}\tilde z_1}$ and within the thickness of $w_{\tt ER}={\textstyle\frac{1}{2}}d_{\tt r}$, formed by the transmitted light. However, the telescope sees a much larger annulus with the thickness of $w_{\tt tel}\simeq z_{\tt r}(\lambda/d_{\tt r})$, which is $w_{\tt tel}/w_{\tt ER}=2z_{\tt r}(\lambda/d^2_{\tt r})=9.72\times 10^7\,(z_{\tt r}/650\,{\rm AU})(\lambda/1\,\mu{\rm m})(1\,{\rm m}/d_{\tt r})^2$ times thicker, thus receiving the light from a much larger area. 

While working on the SGL, we developed an approach to account for the contribution of the solar corona to imaging with the SGL. We take our Sun as the reference luminosity.  
In the region occupied by the Einstein ring in the focal plane of a diffraction-limited telescope, the corona contribution is given after the coronagraph as:
{}
\begin{eqnarray}
P_{\tt cor}(\lambda,\Delta\lambda)
&=&\epsilon_{\tt cor}\,\pi({\textstyle\frac{1}{2}}d_{\tt r})^2
\int_0^{2\pi}\hskip -4pt d\phi
\int_{\theta_0}
^\infty\hskip -4pt \theta d\theta \, B_{\tt cor}(\theta,\lambda,\Delta\lambda),
  \label{eq:pow-fp=+*}
\end{eqnarray}
where $\theta=\rho/ z_{\tt r},$ $ \theta_0=R_\odot/z_{\tt r}$, and $\epsilon_{\tt cor}=0.36$ is the fraction of the encircled energy for the solar corona  (see \cite{Turyshev-Toth:2020-im-extend}).

The surface brightness of the solar corona $B_{\tt cor}$ may be taken from the Baumbach model \citep{Baumbach:1937,Golub-Pasachoff-book:2017}, or use a more recent and a bit more conservative model \cite{November:1996} that is given by:
{}
\begin{align}
B_{\tt cor}(\theta,\lambda,\Delta\lambda)&{}= \Big(10^{-6}\cdot B_\odot (\lambda, \Delta \lambda)\Big)\Big[3.670 \Big(\frac{\theta_0}{\theta}\Big)^{18}+1.939\Big(\frac{\theta_0}{\theta}\Big)^{7.8}+ 5.51\times 10^{-2} \Big(\frac{\theta_0}{\theta}\Big)^{2.5}\Big],
\label{eq:model-th0}
\end{align}
where $B_\odot (\lambda, \Delta \lambda)$ is from (\ref{eq:model-L0*}).  Using (\ref{eq:model-th0}) in (\ref{eq:pow-fp=+*}) and following the approach used in  \cite{Turyshev-Toth:2020-im-extend,Turyshev-Toth:2022a}, we obtain the power of the signal received from the corona and the corresponding photon flux
{}
\begin{eqnarray}
P_{\tt cor}(\lambda,\Delta\lambda)&\simeq&6.58\times 10^{-12}
\Big(\frac{1\,\mu{\rm m}}{\lambda}\Big)^5
\Big(\frac{L_1}{L_\odot}\Big)
\Big(\frac{\Delta \lambda}{10\,{\rm nm}}\Big)
\Big(\frac{d_{\tt r}}{1\,{\rm m}}\Big)^2
\Big(\frac{650\,{\rm AU}}{z_{\tt r}}\Big)^2~~~{\rm W},
\nonumber\\
Q_{\tt cor} =
\frac{\lambda}{hc}P_{\tt cor}&\simeq&
3.31\times 10^7
\Big(\frac{1\,\mu{\rm m}}{\lambda}\Big)^4
\Big(\frac{L_1}{L_\odot}\Big)
\Big(\frac{\Delta \lambda}{10\,{\rm nm}}\Big)
\Big(\frac{d_{\tt r}}{1\,{\rm m}}\Big)^2
\Big(\frac{650\,{\rm AU}}{z_{\tt r}}\Big)^2~~~{\rm phot/s}.
\label{eq:pow-fp=+*4+2}
\end{eqnarray}

Although these estimates may need to be refined by analyzing the spectral composition of the corona of a particular stellar lens and the relevant signal as was recently done in \citep{Turyshev-Toth:2022b}, the obtained results provide good initial estimates.

\subsubsection{Two lenses: transmission and reception from/to the focal regions}
\label{sec:nosie-2-lens}

In the case of a two-lens transmission, the situation changes again. This time, one would have to deal not only with the light received from lens 2 that falls within the annulus around the Einstein ring as in the case discussed in Sec.~\ref{sec:nosie-receiver}, but also the light that arrives into this annulus from lens 1 must also be accounted for. However, in this case, the light from lens 1 will be amplified by the gravitational field of lens 2, providing additional noise background. 

Considering the formation of this background light, we note that this process is very similar to imaging exoplanets with the SGL \cite{Turyshev-Toth:2020-im-extend}. Here, however, we are concerned with the formation of another Einstein ring around lens 2 at the same location as the transmitted signal, produced by light received from lens 1. 

Similar to the SGL, lens 2 will focus light while reducing the size of the image compared to the source by a factor of $z_{\tt r}/z_{12}\sim3.15\times 10^{-4}\,(z_{\tt r}/650 ~{\rm AU}) (10\,{\rm pc}/z_{12})$, see \cite{Turyshev-Toth:2020-im-extend}. For a lens with physical radius $R_1$, positioned at a distance of $z_{12}$ from lens 2, the image of this target at a  distance of $z_{\tt r}$, will be compressed to a cylinder with radius
{}
\begin{equation}
\rho^\star_1=\frac{ z_{\tt r}}{z_{12}}R_1=219.24\,\Big(\frac{R_1}{R_\odot}\Big)\Big(\frac{z_{\tt r}}{650 ~{\rm AU}}\Big) \Big(\frac{10~{\rm pc}}{z_{12}}\Big)~{\rm km}.
\label{eq:rE}
\end{equation}
 Using (\ref{eq:rE}), we see that a telescope with the aperture $d_{\tt r}$ will resolve this object with $N_d$ linear resolution elements,
{}
\begin{equation}
N_d=\frac{2\rho^\star_1}{d_{\tt r}}=\frac{2R_1}{d_{\tt r}}\frac{ z_{\tt r}}{z_{12}}=4.39\times 10^5\,\,\Big(\frac{R_1}{R_\odot}\Big)\Big(\frac{z_{\tt r}}{650 ~{\rm AU}}\Big) \Big(\frac{10~{\rm pc}}{z_{12}}\Big)\Big(\frac{1\,{\rm m}}{d_{\tt r}}\Big).
\label{eq:res-el}
\end{equation}
Equation~(\ref{eq:res-el}) shows that lens 1 is highly resolved in the focal plane for the fiducial parameters.

Therefore, to account for the amplified stellar light from lens 1, we use the expressions developed for the SGL to describe the total power collected from a resolved source.  The relevant expression in this case is Eq.~(2) in \cite{Turyshev-Toth:2022a}.  Let $B_1(\lambda,\Delta\lambda)$ denote the band-integrated photospheric surface brightness of lens 1.  For a blackbody this may be estimated as $B_\lambda(T_1)\Delta\lambda$; if one uses a total band luminosity $L_1(\lambda,\Delta\lambda)$ instead, then $B_1=L_1/(4\pi^2R_1^2)$ must be used.  The power at the Einstein ring in the focal plane of an optical telescope is dominated by the blur and is given as
{}
\begin{eqnarray}
P^{ \star}_{\tt2GL}(\rho)&=&
 \epsilon_{\tt blur}\pi B_1(\lambda,\Delta\lambda)
 \pi({\textstyle\frac{1}{2}}d_{\tt r})^2
 \frac{2R_1}{z_{12}}
 \sqrt{\frac{2r_{g_2}}{ z_{\tt r}}}\,\mu(\rho),
\nonumber\\
&&\hskip 12pt
{\rm with}\qquad
\mu(\rho)=\frac{2}{\pi}{\tt E}\Big[\frac{\rho}{\rho^\star_1}\Big],
\qquad 0\leq \rho/\rho^\star_1\leq 1.
  \label{eq:Pexo-blur}
\end{eqnarray}
In Eq.~(\ref{eq:Pexo-blur}), $\epsilon_{\tt blur}=0.69$ is the encircled energy fraction for the light received from the entire resolved lens 1  \citep{Turyshev-Toth:2020-im-extend} and  ${\tt E}[x]$ is the elliptic integral  \citep{Abramovitz-Stegun:1965}. To evaluate the worst case, below we take $\rho=0$ and thus, $\mu(0)=1$.  

Using $B_\odot(\lambda,\Delta\lambda)$ from (\ref{eq:model-L0*}) as the fiducial normalization, we obtain
{}
\begin{eqnarray}
P^{ \star}_{\tt2GL}
&\simeq&
6.42\times 10^{-9}
\Big(\frac{B_1}{B_\odot}\Big)
\Big(\frac{1\,\mu{\rm m}}{\lambda}\Big)^5
\Big(\frac{\Delta \lambda}{10\,{\rm nm}}\Big)
\Big(\frac{d_{\tt r}}{1~{\rm m}}\Big)^2
\Big(\frac{R_1}{R_\odot}\Big)
\Big(\frac{10~{\rm pc}}{z_{12}}\Big)
\Big(\frac{M_2}{M_\odot}\Big)^\frac{1}{2}
\Big(\frac{650~{\rm AU}}{z_{\tt r}}\Big)^\frac{1}{2}~~~{\rm W},
\label{eq:model-2GL}
\\
Q^{ \star}_{\tt2GL}
=
\frac{\lambda}{hc}P^{\tt \star}_{\tt2GL}
&\simeq&
3.23 \times 10^{10}
\Big(\frac{B_1}{B_\odot}\Big)
\Big(\frac{1\,\mu{\rm m}}{\lambda}\Big)^4
\Big(\frac{\Delta \lambda}{10\,{\rm nm}}\Big)
\Big(\frac{d_{\tt r}}{1~{\rm m}}\Big)^2
\Big(\frac{R_1}{R_\odot}\Big)
\Big(\frac{10~{\rm pc}}{z_{12}}\Big)
\Big(\frac{M_2}{M_\odot}\Big)^\frac{1}{2}
\Big(\frac{650~{\rm AU}}{z_{\tt r}}\Big)^\frac{1}{2}~~~{\rm phot/s}.
\label{eq:model-2GLFF}
\end{eqnarray}

The power in (\ref{eq:model-2GL}) and the photon flux in (\ref{eq:model-2GLFF}) form a challenging background for two-lens power transmissions. 

\subsection{Interstellar transmission: relevant SNRs}
\label{sec:SNRs}

We assume that deterministic contributions from stellar backgrounds can be modeled or subtracted, while their photon shot noise remains.  For photon rates $Q_{\tt sig}$ and $Q_{\tt noise}$ observed for an integration time $t$, the shot-noise-limited SNR is
\begin{align}
{\rm SNR}
=
\frac{Q_{\tt sig}t}{\sqrt{(Q_{\tt sig}+Q_{\tt noise})t}}
=
\frac{Q_{\tt sig}}{\sqrt{Q_{\tt sig}+Q_{\tt noise}}}\sqrt{t}.
\label{eq:snr-def}
\end{align} 

The first case we consider is when the transmitter sends a signal directly via its local lens, with the receiver at an interstellar distance. In this case, using the SNR definition (\ref{eq:snr-def}) and results (\ref{eq:powed-l-Q}) and (\ref{eq:pow-fp==2}), we estimate the relevant SNR as below
{}
\begin{align}
{\rm SNR}^{\tt t}_{\tt1GL}
&=
\frac{Q^{\tt t}_{\tt1GL}}
{\sqrt{Q^{\tt t}_{\tt1GL}+Q^{\tt t\, \star}_{\tt1GL}}}
\sqrt{\frac{t}{1\,{\rm s}}}
\nonumber\\
&\simeq
\frac{1.54\times10^6\,X_{\tt t}}
{\sqrt{1.54\times10^6\,X_{\tt t}+6.78\times10^6\,Y_{\tt t}}}
\sqrt{\frac{t}{1\,{\rm s}}},
\label{eq:snr-cor-t}
\end{align}
where
\begin{align}
X_{\tt t}
&=
\Big(\frac{P_0}{1~{\rm W}}\Big)
\Big(\frac{1~\mu{\rm m}}{\lambda}\Big)^2
\Big(\frac{d_{\tt t}}{1~{\rm m}}\Big)^2
\Big(\frac{d_{\tt r}}{1~{\rm m}}\Big)^2
\Big(\frac{10~{\rm pc}}{z_{\tt r}}\Big)^2
\Big(\frac{M_1}{M_\odot}\Big),
\nonumber\\
Y_{\tt t}
&=
\Big(\frac{L_1}{L_\odot}\Big)
\Big(\frac{1\,\mu{\rm m}}{\lambda}\Big)^4
\Big(\frac{\Delta \lambda}{10\,{\rm nm}}\Big)
\Big(\frac{d_{\tt r}}{1~{\rm m}}\Big)^2
\Big(\frac{10~{\rm pc}}{z_{\tt r}}\Big)^2 .
\end{align}
This represents a high SNR level under the idealized assumptions above, indicating that transmission from a focal region behind a lens to a receiver at an interstellar distance could be practical if the transmitter throughput and background subtraction requirements can be met.

Next, we consider the case when the transmitter sends a signal toward a receiver that uses its local lens to amplify the signal.  In this case, there are two scenarios available: 1)  an isolated transmitter is positioned in space with no significant stellar background, and 2) when the transmitter is positioned next to a star, i.e., a situation similar to transmitting a signal from a ground-based observatory.    

In the first of these two cases, the only background noise to consider is the corona noise from the lensing star next to the receiver.   Using  (\ref{eq:powed-l02-Q}) to represent the signal of interest, $Q^{\tt r}_{\tt1GL}$,  and (\ref{eq:pow-fp=+*4+2}) that for the noise from the stellar corona, $Q_{\tt cor}$, we estimate the SNR for the case when a single lens is positioned at the receiving end of the link: 
{}
\begin{align}
{\rm SNR}^{\tt r\,[1]}_{\tt1GL}&=\frac{Q^{\tt r}_{\tt1GL}}{\sqrt{Q^{\tt r}_{\tt1GL}+Q_{\tt cor}}}\sqrt{\frac{t}{1\,{\rm s}}}\simeq
\nonumber\\&\simeq 
6.97 \,
\Big(\frac{P_0}{1~{\rm W}}\Big)\Big(\frac{\lambda}{1~\mu{\rm m}}\Big)\Big(\frac{10\,{\rm nm}}{\Delta \lambda}\Big)^\frac{1}{2}
\Big(\frac{d_{\tt t}}{1~{\rm m}}\Big)^2
\Big(\frac{10~{\rm pc}}{z_{\tt t}}\Big)^2
\Big(\frac{M_1}{M_\odot}\Big)^\frac{1}{2}
\Big(\frac{z_{\tt r}}{650\,{\rm AU}}\Big)^\frac{3}{2}
\Big(\frac{L_\odot}{L_1}\Big)^\frac{1}{2}
\sqrt{\frac{t}{1\,{\rm s}}}.
\label{eq:snr-cor-r1}
\end{align}

In the second case, we consider the pessimistic geometry in which the transmitter is close to a background star directly behind it.  Let $D$ denote the distance from that background star to the receiving lens, and let $B_\star(\lambda,\Delta\lambda)$ and $R_\star$ denote its band-integrated photospheric surface brightness and radius.  In the regime where the amplified background-star shot noise dominates over the local coronal noise, Eq.~(\ref{eq:model-2GLFF}) may be used with $z_{12}\rightarrow D$ and $M_2\rightarrow M_1$, yielding
{}
\begin{align}
{\rm SNR}^{\tt r\,[2]}_{\tt1GL}
&=
\frac{Q^{\tt r}_{\tt1GL}}
{\sqrt{Q^{\tt r}_{\tt1GL}+Q_{\tt cor}+Q^{\tt \star}_{\tt bg}}}
\sqrt{\frac{t}{1\,{\rm s}}}
\nonumber\\
&\simeq
0.22
\Big(\frac{P_0}{1~{\rm W}}\Big)
\Big(\frac{\lambda}{1~\mu{\rm m}}\Big)
\Big(\frac{10\,{\rm nm}}{\Delta \lambda}\Big)^\frac{1}{2}
\Big(\frac{d_{\tt t}}{1~{\rm m}}\Big)^2
\Big(\frac{10~{\rm pc}}{D}\Big)^\frac{3}{2}
\nonumber\\
&\hskip 22pt \times
\Big(\frac{M_1}{M_\odot}\Big)^\frac{1}{4}
\Big(\frac{z_{\tt r}}{650\,{\rm AU}}\Big)^\frac{3}{4}
\Big(\frac{B_\odot}{B_\star}\Big)^\frac{1}{2}
\Big(\frac{R_\odot}{R_\star}\Big)^\frac{1}{2}
\sqrt{\frac{t}{1\,{\rm s}}}.
\label{eq:snr-cor-r2}
\end{align}

The estimates (\ref{eq:snr-cor-r1}) and (\ref{eq:snr-cor-r2}) suggest that transmission from an interstellar distance directly into the focal region of a lens can be viable under the idealized assumptions used here.  

Finally, we estimate the SNR for the two-lens transmission. In this case, we use  (\ref{eq:powed-l2-Q}) for the signal with  the relevant noise source for the corona around lens 2, $Q_{\tt cor}$, taken from (\ref{eq:pow-fp=+*4+2}) and the amplified light from lens 1, $Q^{\tt \star}_{\tt2GL}$, given by  (\ref{eq:model-2GLFF}). As a result we have the following SNR for the two-lens transmission
{}
\begin{align}
{\rm SNR}_{\tt2GL}
&=
\frac{Q_{\tt2GL}}
{\sqrt{Q_{\tt2GL}+Q_{\tt cor}+Q^{\tt \star}_{\tt2GL}}}
\sqrt{\frac{t}{1\,{\rm s}}}
\nonumber\\
&\simeq
7.05 \times 10^{2}
\Big(\frac{P_0}{1~{\rm W}}\Big)
\Big(\frac{\lambda}{1~\mu{\rm m}}\Big)
\Big(\frac{10\,{\rm nm}}{\Delta \lambda}\Big)^\frac{1}{2}
\Big(\frac{d_{\tt t}}{1~{\rm m}}\Big)^2
\Big(\frac{10~{\rm pc}}{z_{12}}\Big)^\frac{1}{2}
\nonumber\\
&\hskip 48pt \times
\Big(\frac{M_1}{M_\odot}\Big)^\frac{1}{2}
\Big(\frac{M_\odot}{M_2}\Big)^\frac{1}{4}
\Big(\frac{650\,{\rm AU}}{z_{\tt t}}\Big)^\frac{1}{2}
\Big(\frac{z_{\tt r}}{650\,{\rm AU}}\Big)^\frac{1}{4}
\Big(\frac{B_\odot}{B_1}\Big)^\frac{1}{2}
\Big(\frac{R_\odot}{R_1}\Big)^\frac{1}{2}
\sqrt{\frac{t}{1\,{\rm s}}},
\label{eq:snr-2l}
\end{align}
For the fiducial parameters, the two-lens SNR in (\ref{eq:snr-2l}) is the largest of the SNR estimates considered here.  However, the stellar light from lens 1 is also amplified by lens 2, providing a very strong background.  As a result, power transmission with a single lens can yield an SNR competitive with the two-lens transmission.  These SNR estimates should therefore be interpreted as idealized, shot-noise-limited link-budget comparisons rather than as complete engineering demonstrations.

\section{Discussion} 
\label{sec:summary}

Following the motivation outlined in Sec.~\ref{sec:introduction}, we considered the propagation of monochromatic EM waves in the presence of well-separated monopole gravitational lenses in an axially aligned geometry.  For one and two lenses, the resulting diffraction integrals can be evaluated analytically within the thin-lens/eikonal approximation.  In the two-lens case, the second lens produces two closely spaced Einstein-ring contributions that are unresolved by a conventional optical receiver.  The unaveraged PSF depends on both lens masses, but for optical wavelengths and metre-scale receiving apertures the fine structure is averaged over; in that aperture-averaged regime, the final gain is independent of the mass of the second lens. 

The performance of a transmission link is strongly influenced by the factor $z/z_0$ in front of $\vec x'$ in (\ref{eq:amp-Am+}), which determines image scaling in different lensing scenarios: 1) A lens acts as part of a transmitter, when the signal transmission is done from a distance shorter than that of the receiver, $z_0\ll z$. In this case, the image size grows as $z/z_0$, as in the transmitting lens scenario in Sec.~\ref{sec:x-dom} or in the transmission part at lens 1 in the two-lens scenario in  Sec.~\ref{sec:double-C}, when the image size grows as $z_{12}/z_{\tt t}$. 2) A lens acts as part of a receiver with the transmit/receive positions reversed, $z_0\gg z$. In this case, the lens focuses light compressing images by $z/z_0$, as in the receiving-lens diffraction scenario of Sec.~\ref{sec:receiving-lens} and the receiving-link budget of Sec.~\ref{sec:single-B}, or in the reception part of the two-lens transmission  discussed in Sec.~\ref{sec:double-C}.  

This scaling changes the width of the Einstein ring formed at each lens plane and thus the area subtended by the ring, as seen in (\ref{eq:area-ER2}).  Within these approximations, the resulting expressions are useful for order-of-magnitude comparisons among the idealized transmission architectures considered here.  A more complete treatment should include higher-order geometric terms, finite-aperture transmitter fields, non-monopole lens structure, and plasma effects.

Within the idealized link budgets considered here, the two-lens transmission configurations shown by (\ref{eq:powed-l2}) provide the largest light amplification, Eq.~(\ref{eq:mu_av=}), resulting in the gain of $P_{\tt2GL}/P_{\tt free}=\sqrt{{2r_{g_1}}/{z_{\tt t}}}(4z_{12}/d_{\tt r})\simeq 9.62\times 10^{12}\,({M_1}/{M_\odot})^\frac{1}{2}({650\, {\rm AU}}/{z_{\tt t}})^\frac{1}{2}({1~{\rm m}}/{d_{\tt r}})({z_{12}}/{10~{\rm pc}})$. Configurations where a single lens acts as part of a transmitter are the next most favorable, providing access to the light amplification factor of (\ref{eq:amp-PSF1}), thus yielding the gain of  $P^{\tt t}_{\tt1GL}/P_{\tt free}\simeq 2\pi k r_{g_1}\simeq 1.17\times 10^{11} ({M_1}/{M_\odot})({1\,\mu{\rm m}}/{\lambda})$. Finally, configurations where a single lens acts as part of a receiver amplify electromagnetic signals according to (\ref{eq:mu_av}), thus resulting in the gain of $P^{\tt r}_{\tt1GL}/P_{\tt free}= 4 \sqrt{2r_{g_1} z_{\tt r}}/d_{\tt r} \simeq 3.03\times 10^{9}({M_1}/{M_\odot})^\frac{1}{2}({z_{\tt r}}/{650\, {\rm AU}})^\frac{1}{2}({1~{\rm m}}/{d_{\tt r}})$. 

Note that although throughout the paper we used $\lambda=1\mu$m to estimate various values, all our expressions are valid for shorter wavelengths. In fact, not addressing the technical implementations, for shorter wavelength the gravitational lensing will be even more robust with higher overall gains.  Our models work for longer wavelengths as well, but with some limits. As we have shown in \cite{Turyshev-Toth:2019},  longer wavelengths, would experience much higher diffraction, so that one may have to consider using the full diffraction-limited PSFs given by (\ref{eq:S_z*6z-mu2}) and (\ref{eq:amp-psf2-av}). Therefore, these wavelengths were not considered here, as our prior work \cite{Turyshev-Toth:2019,Turyshev-Toth:2019-corona} had shown that such longer wavelengths, say above 30 cm,  will be severely affected by stellar atmospheres to the point of being completely blocked by plasma in their coronas. 

Our key results are (\ref{eq:amp-psf2-av=+}), (\ref{eq:mu_av=}) that present the PSF of a light transmission with a pair of gravitational lenses.  Expression (\ref{eq:amp-psf2-av=+}) has similar mass contributions from the lenses located at both the start and end of the transmission link. The amplification is proportional to the factor $z_{12}/\sqrt{\tilde z_1 \tilde z_2}$, where $z_{12}$ is the distance between the two lenses, while $\tilde z_1$ and $\tilde z_2$ are the effective distances from the transmitter to its respective lens and from the receiver to its lens, respectively. However, as the size of a realistic telescope will be larger than the diffraction pattern set in (\ref{eq:amp-psf2-av=+}), the result must be averaged over the telescope's aperture. We found that after that averaging, the dependence on the mass of the second lens is absent in (\ref{eq:mu_av=}). (This is analogous to (\ref{eq:S_z*6z-mudd+})--(\ref{eq:mu_av}), where the same procedure also removed  a factor of $\sqrt{r_{g_1}}$.) 

For the idealized geometries considered here, the aperture-averaged link budget of a two-lens configuration can exceed that of a single-lens system by a significant margin. We observe that all the transmission cases are characterized by different PSFs, exhibiting different structures with features from a few cm scales to several 100s of meters. Notably, in the transmitting lens case the structure within the PSF (\ref{eq:S_z*6z-mu2+}) is large and does not need to be averaged. However, in the two-lens system the fine structure in the PSF is not going to be directly observed with a reasonable-size telescope. Instead, this structure is averaged out by a modest, say, a 1-m telescope, yielding the average PSFs as in the receiving lens scenario (\ref{eq:mu_av}) and that of the two-lens system (\ref{eq:amp-mu2}). 

The possible feasibility of interstellar transmission is suggested by the SNR estimates for the idealized scenarios considered in Sec.~\ref{sec:SNRs}. Although the two-lens transmission benefits from an impressive combined power amplification, it is also severely affected by the amplified background light coming from the first lens. Other sources of light in the vicinity of the first lens would need to be assessed for specific systems. Both single-lens transmission cases, where the lens is either close to the transmitter or near the receiver, also show robust link performance in the corresponding SNR estimates (\ref{eq:snr-cor-t})--(\ref{eq:snr-cor-r2}). 

Concluding, we note that the results obtained here could be relevant for applications aiming at interstellar power transmission. Not only we can look for transmitted signals using modern astronomical techniques \cite{Clark:2018,Hippke:2018,Kerby-Wright:2021,Tusay:2022,Gillon:2022,Marcy:2022}, we may also transmit such signals with space-based platforms in the focal region of the solar gravitational lens (SGL) using technologies that are either extant or in active development \cite{Helvajian-etal:2022,Turyshev-etal:2023}.

Looking ahead, it would be interesting to explore non-axially symmetric setups and also transmission in the presence of non-spherically symmetric lenses, such as those discussed in \cite{Turyshev-Toth:2021-multipoles,Turyshev-Toth:2021-caustics,Turyshev-Toth:2021-quartic,Turyshev-Toth:2022-STF}.  These lenses will have complex PSF structures exhibiting contributions from various caustics. Although such lenses will disperse some light toward the cusps of the caustics \cite{Turyshev-Toth:2023-faint}, with a proper transmission alignment such effects may be reduced. Investigating such configurations via numerical simulations might shed light on the intricate dynamics in such lensing systems. Nonetheless, our primary focus should be directed towards deriving a formal solution for the situation where a transmitter, positioned at a significant yet finite distance from a lens, emits a beam---whether plane, spherical, or Gaussian---toward that lens. Such a solution may be obtained by following the approach taken in \cite{Turyshev-Toth:2017} that would put all the related analysis on a much stronger footing. Work on this matter is currently in progress, and findings will be reported elsewhere.

\begin{acknowledgments}
We acknowledge discussions with Jason T. Wright who kindly provided us with valuable comments and suggestions on various topics addressed in this document. This work was performed at the Jet Propulsion Laboratory, California Institute of Technology, under a contract with the National Aeronautics and Space Administration. 

\end{acknowledgments}

\appendix
\section{Alternative derivation for a two-lens transmission}
\label{sec:alt-der}

Here we provide an alternative derivation for the diffraction integral (\ref{eq:amp-A2*1}), which is repeated below, for convenience:
{}
\begin{eqnarray}
F_{\tt 2GL}(\vec x) &=&
  \sqrt{2\pi k r_{g_1}}  
 \frac{k}{i \tilde z_{2}}
  \int_0^\infty b_2db_2 \,J_0\big(k\theta_{1}b_2\big)J_0\Big(k\frac{b_2\rho}{  z_{\tt r}}\Big)\exp\Big[ik\Big(\frac{b_2^2}{2 \tilde z_{2}}-2 r_{g_2}\ln(k b_2)\Big)\Big].~~~
  \label{eq:amp-A2*=}
\end{eqnarray}

Clearly, for any impact parameters smaller than the radius of the second lensing star, $b_2\leq R_2$, the light rays will be absorbed by that lens.  Therefore, the integration over $db_2$ in (\ref{eq:amp-A2*=}) really starts from $R_2$. In this case, the argument of the first Bessel function in this expression   is very large and is evaluated to be $k\theta_{1}b_2\simeq kb_2\sqrt{2r_{g_1} \tilde z_{1}}/z_{12}\geq 1.07\times 10^{7}\,(1~\mu{\rm m}/\lambda)(M_1/M_\odot)^\frac{1}{2}(z_{\tt t}/650~{\rm AU})^\frac{1}{2}(10~{\rm pc}/z_{12}) (b_2/R_\odot)$. In this case, this Bessel function, $J_0(x)$ can be approximated by using its expression for large arguments (see \cite{Turyshev-Toth:2020-image})
 {}
\begin{eqnarray}
J_0\big(k\theta_{1}b_2\big)
=\frac{1}{\sqrt{2\pi k\theta_{1}b_2}}\Big(e^{i(k\theta_{1}b_2-\frac{\pi}{4})}+e^{-i(k\theta_{1}b_2-\frac{\pi}{4})}\Big).
  \label{eq:bf0}
\end{eqnarray}
 Note that, as we are concerned with the regions on the optical axis after  lens 2, where $\rho\approx 0$, we do not need to use a similar approximation for the second Bessel function in (\ref{eq:amp-A2*=}). 

As a result of using approximation (\ref{eq:bf0}), expression (\ref{eq:amp-A2*=})  takes the form:
{}
\begin{eqnarray}
F_{\tt 2GL}(\vec x) &=&
\sqrt{2\pi k r_{g_1}} 
\frac{k}{i  \tilde z_{2}} 
 \int_{R_2}^\infty \frac{\sqrt{ b_2}db_2}{\sqrt{2\pi k\theta_{1}}} J_0\Big(k\frac{b_2\rho}{ z_{\tt r}}\Big)
 \Big(e^{i\big(k\theta_{1}b_2-\frac{\pi}{4}\big)}+e^{-i\big(k\theta_{1}b_2-\frac{\pi}{4}\big)}\Big)e^{ik\big(\frac{b_2^2}{2 \tilde z_{2}}-2 r_{g_2}\ln(k b_2)\big)}.~~~
  \label{eq:adph}
\end{eqnarray}

We evaluate this integral using the method of stationary phase (as was done in \cite{Turyshev-Toth:2019-extend}). To do that, we see that the relevant $b_2$-dependent part of the phase in (\ref{eq:adph}) is of the form
{}
\begin{equation}
\varphi_{\pm}(b_2)=k\Big(\frac{b_2^2}{2 \tilde z_{2}} 
\pm b_2\theta_{1}-2 r_{g_2}\ln(k b_2)\Big)\mp \textstyle{\frac{\pi}{4}}.
\label{eq:S-l}
\end{equation}

The phase is stationary when $d\varphi_{\pm}/db_2=0$, which implies
{}
\begin{equation}
\frac{b_2}{ \tilde z_{2}}-\frac{2r_{g_2}}{b_2} \pm\theta_{1}=0.
\label{eq:S-l-pri=}
\end{equation}
As impact parameter is a positive quantity, $b_2\geq 0$, solving (\ref{eq:S-l-pri=}) yields two solutions:
{}
\begin{equation}
b^\pm_2=   \tilde z_{2}{\textstyle\frac{1}{2}}\Big(\sqrt{ \theta_{1}^2+4\theta_2^2} \mp\theta_{1}\Big),
 \qquad {\rm where} \qquad 
 \theta_2=\sqrt{\frac{2r_{g_2}}{\tilde z_{2}}}\simeq 
 \sqrt{\frac{2r_{g_2}}{ z_{\tt r}}}
 \simeq 7.79 \times 10^{-6} \Big(\frac{M_2}{M_\odot}\Big)^\frac{1}{2}\Big(\frac{650\,{\rm AU}}{z_{\tt r}}\Big)^\frac{1}{2}\, {\rm rad},
\label{eq:S-l-pri}
\end{equation}
where  $\theta_2$ is the radius of the Einstein ring to be formed around the second lens. The impact parameters $b^\pm_2$ are two families of impact parameters describing incident and scattered EM waves, corresponding to light rays passing by the near side and the far side of lens 2, correspondingly (see \cite{Turyshev-Toth:2019-extend} for details).

Note that, as opposed to the case with one lens (see (\ref{eq:amp-A1*24}) and (\ref{eq:S_z*6z-mu2})), the two solutions $b^\pm_2$ from (\ref{eq:S-l-pri}) suggest that there are now two Einstein rings formed around the second lens with the radii $\theta^\pm_2=b^\pm_2/\tilde z_{2}$   given as below:
{}
\begin{equation}
\theta^\pm_2= {\textstyle\frac{1}{2}}\Big(\sqrt{\theta_1^2+4\theta_2^2}\mp\theta_1\Big).
\label{eq:S-l-the2pm}
\end{equation}

Note that, given the fact that nominally $\theta^{\tt t}_1\ll \theta_2$ (see (\ref{eq:theta1-t12})) and (\ref{eq:S-l-pri})), both rings are very close being only $\delta \theta_2= \theta^-_2-\theta^+_2 \simeq \theta_1\simeq 2.46 \times 10^{-9}\,{\rm rad}$, which is the angle that is unresolvable by a conventional telescope. So, nominally such a telescope would see both of these rings as one, receiving the signal from both of them. Although these rings may not be resolved from each other, yet they carry the relevant energy to the receiver.

Following the approach presented in  \cite{Turyshev-Toth:2019-extend}, we again use the method of stationary phase (\ref{eq:S-l}) for both solutions in (\ref{eq:S-l-the2pm}). Thus, the amplitude of the EM field on the image plane at the distance of $z_{\tt r}$ from the second lens is given as  
{}
\begin{eqnarray}
F_{\tt 2GL}(\vec x) &= &
\sqrt{2\pi k r_{g_1}}    e^{i\hat \varphi_{2}}
\Big(a_+e^{i\varphi_+}J_0\big(k\theta^+_2\rho\big)+a_-e^{i\varphi_-}J_0\big(k\theta^-_2\rho\big)\Big),~~
  \label{eq:amp-A2*3}
\end{eqnarray}
where  $\hat \varphi_{2}$ is the same as in (\ref{eq:amp-A2*3=}).
Also,  $b^\pm_2$  are from (\ref{eq:S-l-pri}), with $a_\pm$ and $\varphi_\pm$ as below:
{}
\begin{eqnarray}
a^2_\pm(\theta_1,\theta_2)&=&\frac{\Big({\textstyle\frac{1}{2}} \big(\sqrt{1+4\theta_2^2/\theta_1^2}\mp 1\big)\Big)^2}{\sqrt{1+4\theta_2^2/\theta_1^2}},
\label{eq:apm}\\
\varphi_\pm
(\theta_1,\theta_2)&=&k\Big(r_{g_2}\pm {\textstyle\frac{1}{4}} \tilde z_{2} \theta_1^2 \big(\sqrt{ 1+4\theta_2^2/\theta_1^2} \mp  1\big)-2r_{g_2}\ln\Big({\textstyle\frac{1}{2}} k  \tilde z_{2} \theta_1\big(\sqrt{1+4\theta_2^2/\theta_1^2} \mp 1\big)\Big).
\end{eqnarray}

Similarly to (\ref{eq:S_z*6z-mu2}), we use solution (\ref{eq:amp-A2*3}), to determine the PSF of the two-thin-lens system 
{}
\begin{align}
{\tt PSF}_{\tt 2GL} (\vec x) &
=2\pi k r_{g_1}
\Big(a^2_+J^2_0\big(k \theta^+_2 \rho\big)+a^2_-J^2_0\big(k \theta^-_2 \rho\big)+2a_+a_-J_0\big(k \theta^+_2 \rho\big)J_0\big(k \theta^-_2 \rho\big)\sin[\delta\varphi_\pm]\Big).
  \label{eq:amp-psf2}
\end{align}

Considering  the mixed term in (\ref{eq:amp-psf2}), that contains $\sin[\delta\varphi_\pm]$ with $\delta\varphi_\pm$ is given by 
{}
\begin{align}
\delta\varphi_\pm&=\varphi_+-\varphi_-=
k \tilde z_{2} \Big({\textstyle\frac{1}{2}}\theta_1^2 \sqrt{1+4\theta_2^2/\theta_1^2}-\theta_2^2\ln \frac{\sqrt{1+4\theta_2^2/\theta_1^2}-1}{\sqrt{1+4\theta_2^2/\theta_1^2}+1}\Big),
  \label{eq:amp-A2*d4}
\end{align}
we note that the phase difference (\ref{eq:amp-A2*d4}) is, at optical wavelengths, a rapidly oscillating function of $\tilde z_{2}$ that averages to 0. Therefore, the last term in (\ref{eq:amp-psf2}) may be neglected, allowing us to present the averaged PSF as below:
{}
\begin{eqnarray}
\left<{\tt PSF}_{\tt 2GL} (\vec x)\right> &=&2\pi k r_{g_1}
\Big(a^2_+J^2_0\big(k \theta^+_2 \rho\big)+a^2_-J^2_0\big(k \theta^-_2 \rho\big)\Big).
  \label{eq:amp-psf2-av}
\end{eqnarray}
This result yields the maximum amplification factor for the two-lens system which is obtained by setting $\rho=0$:
{}
\begin{eqnarray}
\mu_{\tt 2GL} =\left<{\tt PSF}_{\tt 2GL} ( 0)\right>
&=&2\pi k r_{g_1}
\Big(a^2_++a^2_-\Big).
  \label{eq:amp-A2*4}
\end{eqnarray}

Using expressions for $a_+ $ and $a_-$ from (\ref{eq:apm}) and remembering the single-lens light amplification factor $\mu_{\tt 1GL} = 2\pi k r_{g_1}$ from (\ref{eq:amp-PSF1}) as well as Einstein angles $\theta_1$ and $\theta_2$ from (\ref{eq:theta1-t12}) and (\ref{eq:S-l-pri}), correspondingly, we present result (\ref{eq:amp-A2*4}) as below
{}
\begin{eqnarray}
\mu_{\tt 2GL} 
=\mu_{\tt 1GL}\,\frac{1}{u}\frac{u^2+2}{\sqrt{u^2+4}}, \qquad 
{\rm where} \qquad 
u=\frac{\theta_1}{\theta_2}
\equiv \sqrt{\frac{M_1}{M_2}}\frac{\sqrt{  \tilde z_1 \tilde  z_2}}{ z_{12}}.
  \label{eq:amp-A2*d}
\end{eqnarray}

Considering the case of two lenses of similar masses, with $r_{g_1}\simeq r_{g_2}$ and $\tilde z_1\simeq \tilde z_2\simeq650~{\rm AU}\ll z_{12}=10~{\rm pc}$, we find $u=\theta_1/\theta_2\simeq3.15\times10^{-4}$ and therefore $(u^2+2)/(u\sqrt{u^2+4})\simeq3.17\times10^3$.

We observe from (\ref{eq:amp-A2*d}) that in the case of lenses with uneven masses, the unaveraged light amplification may be larger. Thus, for exotic cases when $\theta _1\gg \theta_2$, the additional amplification factor scales as $(u^2+2)/(u\sqrt{u^2+4})\simeq 1+(\theta_2/\theta_1)^4+{\cal O}((\theta_2/\theta_1)^6)$, thus offering very little additional amplification. For another limiting case, for $\theta_1\ll \theta_2$, the scaling is $(u^2+2)/(u\sqrt{u^2+4})\simeq \theta_2/\theta_1+{\textstyle\frac{3}{8}}(\theta_1/\theta_2)+{\cal O}((\theta_1/\theta_2)^3)$, thus offering a noticeable additional amplification.

To put this in context, we observe that the masses of stars in the solar neighborhood within 30 pc are within the range\footnote{\url{http://www.solstation.com/stars3/100-gs.htm}} of 0.6--2.7~$M_\odot$. Taking the first lens to be our Sun and using (\ref{eq:amp-A2*d}), this range of realistic stellar masses gives additional unaveraged amplification factors of approximately $2.46\times10^3$ and $5.21\times10^3$, respectively. These values apply only to the unresolved, unaveraged PSF; the aperture-averaged gain in (\ref{eq:amp-mu2}) is independent of $M_2$ under condition (\ref{eq:theta1-diff}).

The estimates above are a bit misleading as they pertain only to the case of apertures that are smaller than the first zero of the Bessel functions $J_0\big(k \theta^\pm_2 \rho\big)$ present in (\ref{eq:amp-psf2-av}), namely $\rho^+_{\tt 2GL}=2.40483/(k \theta^+_2)\simeq 4.911$~cm and $\rho^-_{\tt 2GL}=2.40483/(k \theta^-_2)\simeq 4.909$~cm, which is not practical. Given the fact that both of these values are practically the same as in the receiving lens case, the PSF behavior (\ref{eq:S_z*6z-mudd+}) and (\ref{eq:amp-psf2-av}) will be nearly identical (shown in Fig.~\ref{fig:psf-12}), except the latter expression will be $\theta_2/\theta^{\tt t}_1= 1/u \simeq 3.17\times 10^3$ times larger. 

Therefore, for realistic telescopes, in the context of a two-lens system, similarly to (\ref{eq:mu_av}) and (\ref{eq:mu_av=}), we need an aperture average of the PSF, yielding the relevant light amplification factor. After evaluating the requisite integrals  and applying the large argument approximation to the involved Bessel functions, using (\ref{eq:amp-psf2-av}), we obtain:
{}
\begin{eqnarray}
\left<\mu_{\tt 2GL} \right>
&=&
\frac{1}{\pi ({\textstyle\frac{1}{2}}d_{\tt r})^2}
\int_0^{{\textstyle\frac{1}{2}}d_{\tt r}}\int_0^{2\pi}
{\tt PSF}_{\tt 2GL}({\vec x})\,\rho d\rho d\phi
\nonumber\\
&=&
2\pi k r_{g_1}
\Big\{
 a^2_+\Big[J^2_0\big(k \theta^+_2{\textstyle\frac{1}{2}}d_{\tt r}\big)+J^2_1\big(k \theta^+_2{\textstyle\frac{1}{2}}d_{\tt r}\big)\Big]
+a^2_-\Big[J^2_0\big(k \theta^-_2{\textstyle\frac{1}{2}}d_{\tt r}\big)+J^2_1\big(k \theta^-_2{\textstyle\frac{1}{2}}d_{\tt r}\big)\Big]
\Big\}
\nonumber\\
&\simeq&
\frac{8r_{g_1}}{d_{\tt r}}
\Big(\frac{a^2_+}{\theta^+_2}+\frac{a^2_-}{\theta^-_2}\Big)
=
\sqrt{\frac{2r_{g_1}}{z_{\tt t}}}\frac{4z_{12}}{d_{\tt r}}
=
\left.\left<\mu^{\tt r}_{\tt 1GL}\right>\right|_{z_{\tt r}=z_{\tt t}}
\frac{z_{12}}{z_{\tt t}}
\nonumber\\
&=&
9.62\times 10^{12}
\Big(\frac{M_1}{M_\odot}\Big)^\frac{1}{2}
\Big(\frac{650\, {\rm AU}}{z_{\tt t}}\Big)^\frac{1}{2}
\Big(\frac{1~{\rm m}}{d_{\tt r}}\Big)
\Big(\frac{z_{12}}{10\,{\rm pc}}\Big),
\label{eq:amp-mu2}
\end{eqnarray}
where  $\theta^\pm_2$ and $a^2_\pm$ are given by (\ref{eq:S-l-the2pm}) and (\ref{eq:apm}), correspondingly.  Clearly, this result is identical to (\ref{eq:mu_av=}). 

We see again that after averaging, the PSF (\ref{eq:amp-mu2}) does not depend on the mass of lens 2 in the aperture-averaged regime.  This is because the size of the telescope is larger than the first zero of diffraction pattern in (\ref{eq:amp-psf2-av}), thus satisfying the condition (\ref{eq:theta1-diff}). Note that other parameter choices, especially in the scenarios where the lens is a part of a transmitter, which is characterized by a much larger scale of the diffraction pattern imprinted in the PSF, like in (\ref{eq:S_z*6z-mu2+})--(\ref{eq:theta1-t}), does not require the averaging.

\section{EM propagation in the presence of two lenses}
\label{sec:path-int}

In Sec.~\ref{sec:basics} we discussed diffraction of light in the presence of one lens. Here we show how the same approach can be extended to two and more lenses. Clearly, the amplitude of the EM field  is given by the same equation  (\ref{eq:amp-FK}) where for each lens $i=1,2$, we will have a different impact parameter $b_i$ and the gravitational phase shift, $\psi(\vec b_i)$. 

\begin{figure}
\vspace{-10pt}
  \begin{center}
\includegraphics[width=0.65\linewidth]{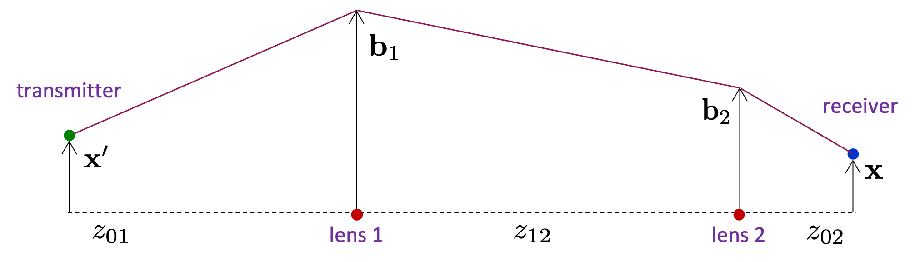}
  \end{center}
  \vspace{-15pt}
  \caption{A two-lens geometry for power transmission via gravitational lensing showing the transmitter, two lenses, and the receiver (compare to Fig.~\ref{fig:geom} for a one-lens transmission geometry). Also shown is the distance from   the transmitter plane to lens 1, $z_{01}$, distance between the lenses,  $z_{12}$, as well as that from lens 2 to the receiver plane, $z_{02}$.  
  }
\label{fig:geom2}
\end{figure}

Using the eikonal (\ref{eq:amp-S+}), the main text specializes to the aligned case. In the case of two lenses, the effective path length (eikonal) $S(\vec x', \vec x, \vec b_1,\vec b_2)$ along a path from a source's position $(\vec x',-z_{01})$ to the observer position $(\vec x,z_{12}+z_{02})$ via points $(\vec b_1,0)$ and $(\vec b_2,z_{12})$ on the lens planes, neglecting terms of order ${\cal O}(b_1^4/z_{01}^3,b_1^4/z_{12}^3,b_2^4/z_{12}^3,b_2^4/z_{02}^3)$, has the form (Fig.~\ref{fig:geom2} shows overall geometry of the gravitational lensing system):
 {}
\begin{eqnarray}
S(\vec x', \vec b_1,  \vec b_2, \vec x) &=&
 \sqrt{(\vec b_1-\vec x')^2+z_{01}^2}+ \sqrt{(\vec b_1-\vec b_2)^2+z^2_{12}}+ \sqrt{(\vec b_2-\vec x)^2+z_{02}^2}-\psi(\vec b_1)-\psi(\vec b_2)= \nonumber\\
 &&\hskip -100pt =\,
 z_{01}+z_{12}+z_{02}+\frac{(\vec b_2-\vec x')^2}{2(z_{01}+z_{12})}+\frac{z_{01}+z_{12}}{2z_{01}z_{12}}\Big(\vec b_1-\frac{z_{01}}{z_{01}+z_{12}}\vec b_2-\frac{z_{12}}{z_{01}+z_{12}}\vec x'\Big)^2+\frac{(\vec b_2-\vec x)^2}{2z_{02}}-\psi(\vec b_1)-\psi(\vec b_2).~~
  \label{eq:amp-S+}
\end{eqnarray}

As a result, the wave amplitude on the observer plane is written as below
{}
\begin{eqnarray}
A(\vec x', \vec x)&=& A_0(\vec x', \vec x)F_{\tt 2GL}(\vec x', \vec x),
  \label{eq:amp00=}
\end{eqnarray}
where, for the axial normalization used in the main text, $A_0$ factors out the axial free propagation phase,
{}
\begin{eqnarray}
A_0(\vec x', \vec x)&=&
\frac{E_0}{(z_{01}+z_{12}+z_{02})}
\exp\Big[ik\big(z_{01}+z_{12}+z_{02}\big)\Big].
\label{eq:amp=}
\end{eqnarray}

With the normalization (\ref{eq:amp=}), a fully off-axis amplification normalized to the direct free-space path would instead factor out the transverse free-space phase $k(\vec x-\vec x')^2/[2(z_{01}+z_{12}+z_{02})]$ and subtract it from the Fermat potential.

In the case of a pair of isolated monopole gravitational lenses, the amplification factor $F_{\tt 2GL}(\vec x', \vec x)$ is given by the following nested diffraction integral
{}
\begin{eqnarray}
F_{\tt 2GL}(\vec x',\vec x) &=&
 \frac{(z_{01}+z_{12}+z_{02})}{(z_{01}+z_{12})z_{02}}\frac{ke^{i\phi_{\tt G2}}}{2\pi i}\iint d^2\vec b_2 \exp\Big[ik\Big(\frac{(\vec b_2-\vec x')^2}{2(z_{01}+z_{12})}+\frac{(\vec b_2-\vec x)^2}{2z_{02}}-2r_{g_2}\ln(k b_2)\Big)\Big]\times
\nonumber\\
 && \hskip -30pt 
 \times ~ \frac{(z_{01}+z_{12})}{z_{01}z_{12}} \frac{ke^{i\phi_{\tt G1}}}{2\pi i}\iint d^2\vec b_1
 \,\hat A_0(\vec b_1)\exp\Big[ik\Big(\frac{z_{01}+z_{12}}{2z_{01}z_{12}}\Big(\vec b_1-\frac{z_{01}}{z_{01}+z_{12}}\Big(\vec b_2+\frac{z_{12}}{z_{01}}\vec x'\Big)\Big)^2-2r_{g_1}\ln(k b_1)\Big)\Big],~~~~
  \label{eq:amp-Am=}
\end{eqnarray}
where the gravitational phase factors are given as $\phi_{\tt G1}=kr_{g_1}\ln(4k^2z_{01}z_{12})$ and $\phi_{\tt G2}=kr_{g_2}\ln(4k^2z_{12}z_{02})$, and $\hat A_0(\vec b_1)$ denotes the source or aperture illumination factor on the first lens plane. 

Result (\ref{eq:amp-Am=}) summarizes the logic of the procedure used in this paper.  It has the non-vanishing position of a transmitter $\vec x' \not=0$.  We take these integrals one after another, each time removing the appropriate spherical waves from the integrand. The integral of (\ref{eq:amp-Am=}) represents the structure of the amplitude of the lensed EM wave that was obtained using the scalar theory of light diffraction using the Fresnel--Kirchhoff diffraction formula \cite{Born-Wolf:1999} and has the form of a path integral  \cite{Feynman-Hibbs:1991} (see \cite{Turyshev-Toth:2021-multipoles} for details.) This logic may be extended to the case involving a larger number of lenses, $i>2$. Furthermore, the same approach will work for extended lenses (i.e., with higher-order multipole moments) with the gravitational shift $\psi(\vec b_i)$  containing all the relevant information about lens' mass distribution.


\begin{thebibliography}{39}%
\makeatletter
\providecommand \@ifxundefined [1]{%
 \@ifx{#1\undefined}
}%
\providecommand \@ifnum [1]{%
 \ifnum #1\expandafter \@firstoftwo
 \else \expandafter \@secondoftwo
 \fi
}%
\providecommand \@ifx [1]{%
 \ifx #1\expandafter \@firstoftwo
 \else \expandafter \@secondoftwo
 \fi
}%
\providecommand \natexlab [1]{#1}%
\providecommand \enquote  [1]{``#1''}%
\providecommand \bibnamefont  [1]{#1}%
\providecommand \bibfnamefont [1]{#1}%
\providecommand \citenamefont [1]{#1}%
\providecommand \href@noop [0]{\@secondoftwo}%
\providecommand \href [0]{\begingroup \@sanitize@url \@href}%
\providecommand \@href[1]{\@@startlink{#1}\@@href}%
\providecommand \@@href[1]{\endgroup#1\@@endlink}%
\providecommand \@sanitize@url [0]{\catcode `\\12\catcode `\$12\catcode
  `\&12\catcode `\#12\catcode `\^12\catcode `\_12\catcode `\%12\relax}%
\providecommand \@@startlink[1]{}%
\providecommand \@@endlink[0]{}%
\providecommand \url  [0]{\begingroup\@sanitize@url \@url }%
\providecommand \@url [1]{\endgroup\@href {#1}{\urlprefix }}%
\providecommand \urlprefix  [0]{URL }%
\providecommand \Eprint [0]{\href }%
\providecommand \doibase [0]{https://doi.org/}%
\providecommand \selectlanguage [0]{\@gobble}%
\providecommand \bibinfo  [0]{\@secondoftwo}%
\providecommand \bibfield  [0]{\@secondoftwo}%
\providecommand \translation [1]{[#1]}%
\providecommand \BibitemOpen [0]{}%
\providecommand \bibitemStop [0]{}%
\providecommand \bibitemNoStop [0]{.\EOS\space}%
\providecommand \EOS [0]{\spacefactor3000\relax}%
\providecommand \BibitemShut  [1]{\csname bibitem#1\endcsname}%
\let\auto@bib@innerbib\@empty
\bibitem [{\citenamefont {{Turyshev}}(2017)}]{Turyshev:2017}%
  \BibitemOpen
  \bibfield  {author} {\bibinfo {author} {\bibfnamefont {S.~G.}\ \bibnamefont
  {{Turyshev}}},\ }\bibfield  {title} {\bibinfo {title} {Wave-theoretical
  description of the solar gravitational lens},\ }\href@noop {} {\bibfield
  {journal} {\bibinfo  {journal} {Phys. Rev. D}\ }\textbf {\bibinfo {volume}
  {95}},\ \bibinfo {pages} {084041} (\bibinfo {year} {2017})},\ \Eprint
  {https://arxiv.org/abs/arXiv:1703.05783 [gr-qc]} {arXiv:arXiv:1703.05783
  [gr-qc]} \BibitemShut {NoStop}%
\bibitem [{\citenamefont {{Turyshev}}\ and\ \citenamefont
  {{Toth}}(2017)}]{Turyshev-Toth:2017}%
  \BibitemOpen
  \bibfield  {author} {\bibinfo {author} {\bibfnamefont {S.~G.}\ \bibnamefont
  {{Turyshev}}}\ and\ \bibinfo {author} {\bibfnamefont {V.~T.}\ \bibnamefont
  {{Toth}}},\ }\bibfield  {title} {\bibinfo {title} {Diffraction of
  electromagnetic waves in the gravitational field of the sun},\ }\href@noop {}
  {\bibfield  {journal} {\bibinfo  {journal} {Phys. Rev. D}\ }\textbf {\bibinfo
  {volume} {96}},\ \bibinfo {pages} {024008} (\bibinfo {year} {2017})},\
  \Eprint {https://arxiv.org/abs/arXiv:1704.06824 [gr-qc]} {arXiv:1704.06824
  [gr-qc]} \BibitemShut {NoStop}%
\bibitem [{\citenamefont {Turyshev}\ and\ \citenamefont
  {Toth}(2022{\natexlab{a}})}]{Turyshev-Toth:2022a}%
  \BibitemOpen
  \bibfield  {author} {\bibinfo {author} {\bibfnamefont {S.~G.}\ \bibnamefont
  {Turyshev}}\ and\ \bibinfo {author} {\bibfnamefont {V.~T.}\ \bibnamefont
  {Toth}},\ }\bibfield  {title} {\bibinfo {title} {{Resolved imaging of
  exoplanets with the solar gravitational lens}},\ }\href
  {https://doi.org/10.1093/mnras/stac2130} {\bibfield  {journal} {\bibinfo
  {journal} {MNRAS}\ }\textbf {\bibinfo {volume} {515}},\ \bibinfo {pages}
  {6122} (\bibinfo {year} {2022}{\natexlab{a}})},\ \Eprint
  {https://arxiv.org/abs/{arXiv:2204.04866 [gr-qc]}} {{arXiv:2204.04866
  [gr-qc]}} \BibitemShut {NoStop}%
\bibitem [{\citenamefont {Turyshev}\ and\ \citenamefont
  {Toth}(2022{\natexlab{b}})}]{Turyshev-Toth:2022b}%
  \BibitemOpen
  \bibfield  {author} {\bibinfo {author} {\bibfnamefont {S.~G.}\ \bibnamefont
  {Turyshev}}\ and\ \bibinfo {author} {\bibfnamefont {V.~T.}\ \bibnamefont
  {Toth}},\ }\bibfield  {title} {\bibinfo {title} {{Spectrally resolved imaging
  with the solar gravitational lens}},\ }\href
  {https://doi.org/10.1103/PhysRevD.106.044059} {\bibfield  {journal} {\bibinfo
   {journal} {Phys. Rev. D}\ }\textbf {\bibinfo {volume} {106}},\ \bibinfo
  {pages} {044059} (\bibinfo {year} {2022}{\natexlab{b}})},\ \Eprint
  {https://arxiv.org/abs/{arXiv:2206.03037 [gr-qc]}} {{arXiv:2206.03037
  [gr-qc]}} \BibitemShut {NoStop}%
\bibitem [{\citenamefont {Eshleman}(1979)}]{vonEshleman:1979}%
  \BibitemOpen
  \bibfield  {author} {\bibinfo {author} {\bibfnamefont {V.~R.}\ \bibnamefont
  {Eshleman}},\ }\bibfield  {title} {\bibinfo {title} {Gravitational lens of
  the sun: Its potential for observations and communications over interstellar
  distances},\ }\href {https://doi.org/10.1126/science.205.4411.1133}
  {\bibfield  {journal} {\bibinfo  {journal} {Science}\ }\textbf {\bibinfo
  {volume} {205}},\ \bibinfo {pages} {1133} (\bibinfo {year}
  {1979})}\BibitemShut {NoStop}%
\bibitem [{\citenamefont {{Maccone}}(2009)}]{Maccone-book:2009}%
  \BibitemOpen
  \bibfield  {author} {\bibinfo {author} {\bibfnamefont {C.}~\bibnamefont
  {{Maccone}}},\ }\href@noop {} {\emph {\bibinfo {title} {Deep Space Flight and
  Communications Exploiting the Sun as a Gravitational Lens}}}\ (\bibinfo
  {publisher} {Springer-Verlag Berlin Heidelberg},\ \bibinfo {year}
  {2009})\BibitemShut {NoStop}%
\bibitem [{\citenamefont {{Maccone}}(2011)}]{Maccone:2011}%
  \BibitemOpen
  \bibfield  {author} {\bibinfo {author} {\bibfnamefont {C.}~\bibnamefont
  {{Maccone}}},\ }\bibfield  {title} {\bibinfo {title} {{Interstellar radio
  links enhanced by exploiting the Sun as a Gravitational Lens}},\ }\href@noop
  {} {\bibfield  {journal} {\bibinfo  {journal} {Acta Astronautica}\ }\textbf
  {\bibinfo {volume} {68}},\ \bibinfo {pages} {76} (\bibinfo {year}
  {2011})}\BibitemShut {NoStop}%
\bibitem [{\citenamefont {{Clark}}\ and\ \citenamefont
  {{Cahoy}}(2018)}]{Clark:2018}%
  \BibitemOpen
  \bibfield  {author} {\bibinfo {author} {\bibfnamefont {J.~R.}\ \bibnamefont
  {{Clark}}}\ and\ \bibinfo {author} {\bibfnamefont {K.}~\bibnamefont
  {{Cahoy}}},\ }\bibfield  {title} {\bibinfo {title} {{Optical Detection of
  Lasers with Near-term Technology at Interstellar Distances}},\ }\href
  {https://doi.org/10.3847/1538-4357/aae380} {\bibfield  {journal} {\bibinfo
  {journal} {ApJ}\ }\textbf {\bibinfo {volume} {867}},\ \bibinfo {eid} {97}
  (\bibinfo {year} {2018})}\BibitemShut {NoStop}%
\bibitem [{\citenamefont {{Kerby}}\ and\ \citenamefont
  {{Wright}}(2021)}]{Kerby-Wright:2021}%
  \BibitemOpen
  \bibfield  {author} {\bibinfo {author} {\bibfnamefont {S.}~\bibnamefont
  {{Kerby}}}\ and\ \bibinfo {author} {\bibfnamefont {J.~T.}\ \bibnamefont
  {{Wright}}},\ }\bibfield  {title} {\bibinfo {title} {{Stellar Gravitational
  Lens Engineering for Interstellar Communication and Artifact SETI}},\
  }\href@noop {} {\bibfield  {journal} {\bibinfo  {journal} {AJ}\ }\textbf
  {\bibinfo {volume} {162}},\ \bibinfo {eid} {252} (\bibinfo {year}
  {2021})}\BibitemShut {NoStop}%
\bibitem [{\citenamefont {{Tusay}}\ \emph {et~al.}(2022)\citenamefont
  {{Tusay}}, \citenamefont {{Huston}}, \citenamefont {{Dedrick}}, \citenamefont
  {{Kerby}}, \citenamefont {{Palumbo}}, \citenamefont {{Croft}}, \citenamefont
  {{Wright}}, \citenamefont {{Robertson}}, \citenamefont {{Sheikh}},
  \citenamefont {{Duffy}}, \citenamefont {{Foote}}, \citenamefont {{Hyde}},
  \citenamefont {{Lafond}}, \citenamefont {{Mullikin}}, \citenamefont
  {{Parts}}, \citenamefont {{Sandhaus}}, \citenamefont {{Smith}}, \citenamefont
  {{Sneed}}, \citenamefont {{Czech}}, \citenamefont {{Gajjar}},\ and\
  \citenamefont {{Breakthrough Listen}}}]{Tusay:2022}%
  \BibitemOpen
  \bibfield  {author} {\bibinfo {author} {\bibfnamefont {N.}~\bibnamefont
  {{Tusay}}}, \bibinfo {author} {\bibfnamefont {M.~J.}\ \bibnamefont
  {{Huston}}}, \bibinfo {author} {\bibfnamefont {C.~M.}\ \bibnamefont
  {{Dedrick}}}, \bibinfo {author} {\bibfnamefont {S.}~\bibnamefont {{Kerby}}},
  \bibinfo {author} {\bibfnamefont {M.~L.}\ \bibnamefont {{Palumbo}}}, \bibinfo
  {author} {\bibfnamefont {S.}~\bibnamefont {{Croft}}}, \bibinfo {author}
  {\bibfnamefont {J.~T.}\ \bibnamefont {{Wright}}}, \bibinfo {author}
  {\bibfnamefont {P.}~\bibnamefont {{Robertson}}}, \bibinfo {author}
  {\bibfnamefont {S.}~\bibnamefont {{Sheikh}}}, \bibinfo {author}
  {\bibfnamefont {L.}~\bibnamefont {{Duffy}}}, \bibinfo {author} {\bibfnamefont
  {G.}~\bibnamefont {{Foote}}}, \bibinfo {author} {\bibfnamefont
  {A.}~\bibnamefont {{Hyde}}}, \bibinfo {author} {\bibfnamefont
  {J.}~\bibnamefont {{Lafond}}}, \bibinfo {author} {\bibfnamefont
  {E.}~\bibnamefont {{Mullikin}}}, \bibinfo {author} {\bibfnamefont
  {W.}~\bibnamefont {{Parts}}}, \bibinfo {author} {\bibfnamefont
  {P.}~\bibnamefont {{Sandhaus}}}, \bibinfo {author} {\bibfnamefont {H.~H.}\
  \bibnamefont {{Smith}}}, \bibinfo {author} {\bibfnamefont {E.~L.}\
  \bibnamefont {{Sneed}}}, \bibinfo {author} {\bibfnamefont {D.}~\bibnamefont
  {{Czech}}}, \bibinfo {author} {\bibfnamefont {V.}~\bibnamefont {{Gajjar}}},\
  and\ \bibinfo {author} {\bibnamefont {{Breakthrough Listen}}},\ }\bibfield
  {title} {\bibinfo {title} {{A Search for Radio Technosignatures at the Solar
  Gravitational Lens Targeting Alpha Centauri}},\ }\href
  {https://doi.org/10.3847/1538-3881/ac8358} {\bibfield  {journal} {\bibinfo
  {journal} {AJ}\ }\textbf {\bibinfo {volume} {164}},\ \bibinfo {eid} {116}
  (\bibinfo {year} {2022})}\BibitemShut {NoStop}%
\bibitem [{\citenamefont {{Gillon}}\ \emph {et~al.}(2022)\citenamefont
  {{Gillon}}, \citenamefont {{Burdanov}},\ and\ \citenamefont
  {{Wright}}}]{Gillon:2022}%
  \BibitemOpen
  \bibfield  {author} {\bibinfo {author} {\bibfnamefont {M.}~\bibnamefont
  {{Gillon}}}, \bibinfo {author} {\bibfnamefont {A.}~\bibnamefont
  {{Burdanov}}},\ and\ \bibinfo {author} {\bibfnamefont {J.~T.}\ \bibnamefont
  {{Wright}}},\ }\bibfield  {title} {\bibinfo {title} {{Search for an Alien
  Message to a Nearby Star}},\ }\href@noop {} {\bibfield  {journal} {\bibinfo
  {journal} {AJ}\ }\textbf {\bibinfo {volume} {164}},\ \bibinfo {eid} {221}
  (\bibinfo {year} {2022})}\BibitemShut {NoStop}%
\bibitem [{\citenamefont {{Turyshev}}\ and\ \citenamefont
  {{Toth}}(2021)}]{Turyshev-Toth:2021-multipoles}%
  \BibitemOpen
  \bibfield  {author} {\bibinfo {author} {\bibfnamefont {S.~G.}\ \bibnamefont
  {{Turyshev}}}\ and\ \bibinfo {author} {\bibfnamefont {V.~T.}\ \bibnamefont
  {{Toth}}},\ }\bibfield  {title} {\bibinfo {title} {Diffraction of
  electromagnetic waves by an extended gravitational lens},\ }\href@noop {}
  {\bibfield  {journal} {\bibinfo  {journal} {Phys. Rev. D}\ }\textbf {\bibinfo
  {volume} {103}},\ \bibinfo {pages} {064076} (\bibinfo {year} {2021})},\
  \bibinfo {note} {arXiv:2102.03891 [gr-qc]}\BibitemShut {NoStop}%
\bibitem [{\citenamefont {Turyshev}\ and\ \citenamefont
  {Toth}(2021)}]{Turyshev-Toth:2021-all-regions}%
  \BibitemOpen
  \bibfield  {author} {\bibinfo {author} {\bibfnamefont {S.~G.}\ \bibnamefont
  {Turyshev}}\ and\ \bibinfo {author} {\bibfnamefont {V.~T.}\ \bibnamefont
  {Toth}},\ }\bibfield  {title} {\bibinfo {title} {Gravitational lensing by an
  extended mass distribution},\ }\href@noop {} {\bibfield  {journal} {\bibinfo
  {journal} {Phys. Rev. D}\ }\textbf {\bibinfo {volume} {104}},\ \bibinfo
  {pages} {044013} (\bibinfo {year} {2021})},\ \bibinfo {note}
  {arXiv:2106.06696 [gr-qc]}\BibitemShut {NoStop}%
\bibitem [{\citenamefont {{Turyshev}}\ and\ \citenamefont
  {{Toth}}(2020{\natexlab{a}})}]{Turyshev-Toth:2020-im-extend}%
  \BibitemOpen
  \bibfield  {author} {\bibinfo {author} {\bibfnamefont {S.~G.}\ \bibnamefont
  {{Turyshev}}}\ and\ \bibinfo {author} {\bibfnamefont {V.~T.}\ \bibnamefont
  {{Toth}}},\ }\bibfield  {title} {\bibinfo {title} {Image formation for
  extended sources with the solar gravitational lens},\ }\href@noop {}
  {\bibfield  {journal} {\bibinfo  {journal} {Phys. Rev. D}\ }\textbf {\bibinfo
  {volume} {102}},\ \bibinfo {pages} {024038} (\bibinfo {year}
  {2020}{\natexlab{a}})},\ \bibinfo {note} {arXiv:2002.06492
  [astro-ph.IM]}\BibitemShut {NoStop}%
\bibitem [{\citenamefont {Herlt}\ and\ \citenamefont
  {Stephani}(1976)}]{Herlt-Stephani:1976}%
  \BibitemOpen
  \bibfield  {author} {\bibinfo {author} {\bibfnamefont {E.}~\bibnamefont
  {Herlt}}\ and\ \bibinfo {author} {\bibfnamefont {H.}~\bibnamefont
  {Stephani}},\ }\bibfield  {title} {\bibinfo {title} {Wave optics of the
  spherical gravitational lens part i: Diffraction of a plane electromagnetic
  wave by a large star},\ }\href {https://doi.org/10.1007/BF01807086}
  {\bibfield  {journal} {\bibinfo  {journal} {Int. J. Theor. Phys.}\ }\textbf
  {\bibinfo {volume} {15}},\ \bibinfo {pages} {45} (\bibinfo {year}
  {1976})}\BibitemShut {NoStop}%
\bibitem [{\citenamefont {{Born}}\ and\ \citenamefont
  {{Wolf}}(1999)}]{Born-Wolf:1999}%
  \BibitemOpen
  \bibfield  {author} {\bibinfo {author} {\bibfnamefont {M.}~\bibnamefont
  {{Born}}}\ and\ \bibinfo {author} {\bibfnamefont {E.}~\bibnamefont
  {{Wolf}}},\ }\href@noop {} {\emph {\bibinfo {title} {Principles of Optics:
  Electromagnetic Theory of Propagation, Interference and Diffraction of
  Light}}}\ (\bibinfo  {publisher} {Cambridge University Press; 7th edition},\
  \bibinfo {year} {October 13, 1999})\BibitemShut {NoStop}%
\bibitem [{\citenamefont {{Landau}}\ and\ \citenamefont
  {{Lifshitz}}(1988)}]{Landau-Lifshitz:1988}%
  \BibitemOpen
  \bibfield  {author} {\bibinfo {author} {\bibfnamefont {L.~D.}\ \bibnamefont
  {{Landau}}}\ and\ \bibinfo {author} {\bibfnamefont {E.~M.}\ \bibnamefont
  {{Lifshitz}}},\ }\href@noop {} {\emph {\bibinfo {title} {The Classical Theory
  of Fields.}}}\ (\bibinfo  {publisher} {7th edition. Nauka: Moscow (in
  Russian)},\ \bibinfo {year} {1988})\BibitemShut {NoStop}%
\bibitem [{\citenamefont {{Turyshev}}\ and\ \citenamefont
  {{Toth}}(2022)}]{Turyshev-Toth:2022-STF}%
  \BibitemOpen
  \bibfield  {author} {\bibinfo {author} {\bibfnamefont {S.~G.}\ \bibnamefont
  {{Turyshev}}}\ and\ \bibinfo {author} {\bibfnamefont {V.~T.}\ \bibnamefont
  {{Toth}}},\ }\bibfield  {title} {\bibinfo {title} {{Multipole decomposition
  of gravitational lensing}},\ }\href
  {https://doi.org/10.1103/PhysRevD.105.024022} {\bibfield  {journal} {\bibinfo
   {journal} {Phys. Rev. D}\ }\textbf {\bibinfo {volume} {105}},\ \bibinfo
  {eid} {024022} (\bibinfo {year} {2022})},\ \Eprint
  {https://arxiv.org/abs/arXiv:2107.13126 [gr-qc]} {arXiv:arXiv:2107.13126
  [gr-qc]} \BibitemShut {NoStop}%
\bibitem [{\citenamefont {{Turyshev}}\ and\ \citenamefont
  {{Toth}}(2019)}]{Turyshev-Toth:2019-extend}%
  \BibitemOpen
  \bibfield  {author} {\bibinfo {author} {\bibfnamefont {S.~G.}\ \bibnamefont
  {{Turyshev}}}\ and\ \bibinfo {author} {\bibfnamefont {V.~T.}\ \bibnamefont
  {{Toth}}},\ }\bibfield  {title} {\bibinfo {title} {Imaging extended sources
  with the solar gravitational lens},\ }\href
  {https://doi.org/10.1103/PhysRevD.89.105029} {\bibfield  {journal} {\bibinfo
  {journal} {Phys. Rev. D}\ }\textbf {\bibinfo {volume} {100}},\ \bibinfo
  {pages} {084018} (\bibinfo {year} {2019})},\ \bibinfo {note}
  {arXiv:1908.01948 [gr-qc]}\BibitemShut {NoStop}%
\bibitem [{\citenamefont {{Turyshev}}\ and\ \citenamefont
  {{Toth}}(2018{\natexlab{a}})}]{Turyshev-Toth:2018}%
  \BibitemOpen
  \bibfield  {author} {\bibinfo {author} {\bibfnamefont {S.~G.}\ \bibnamefont
  {{Turyshev}}}\ and\ \bibinfo {author} {\bibfnamefont {V.~T.}\ \bibnamefont
  {{Toth}}},\ }\bibfield  {title} {\bibinfo {title} {Wave-optical treatment of
  the shadow cast by a large sphere},\ }\href@noop {} {\bibfield  {journal}
  {\bibinfo  {journal} {Phys. Rev. A}\ }\textbf {\bibinfo {volume} {97}},\
  \bibinfo {pages} {033810} (\bibinfo {year} {2018}{\natexlab{a}})},\ \Eprint
  {https://arxiv.org/abs/arXiv:1801.06253 [physics.optics]} {arXiv:1801.06253
  [physics.optics]} \BibitemShut {NoStop}%
\bibitem [{\citenamefont {{Turyshev}}\ and\ \citenamefont
  {{Toth}}(2018{\natexlab{b}})}]{Turyshev-Toth:2018-grav-shadow}%
  \BibitemOpen
  \bibfield  {author} {\bibinfo {author} {\bibfnamefont {S.~G.}\ \bibnamefont
  {{Turyshev}}}\ and\ \bibinfo {author} {\bibfnamefont {V.~T.}\ \bibnamefont
  {{Toth}}},\ }\bibfield  {title} {\bibinfo {title} {Wave-optical treatment of
  the shadow cast by a large gravitating sphere},\ }\href@noop {} {\bibfield
  {journal} {\bibinfo  {journal} {Phys. Rev. D}\ }\textbf {\bibinfo {volume}
  {98}},\ \bibinfo {pages} {104015} (\bibinfo {year} {2018}{\natexlab{b}})},\
  \bibinfo {note} {arXiv:1805.10581 [gr-qc]}\BibitemShut {NoStop}%
\bibitem [{\citenamefont {{Wolf}}(1959)}]{Wolf-Gabor:1959}%
  \BibitemOpen
  \bibfield  {author} {\bibinfo {author} {\bibfnamefont {E.}~\bibnamefont
  {{Wolf}}},\ }\bibfield  {title} {\bibinfo {title} {Electromagnetic
  diffraction in optical systems - i. an integral representation of the image
  field},\ }\href@noop {} {\bibfield  {journal} {\bibinfo  {journal} {Proc.
  Royal Soc. London. Series A. Math. Phys. Sci.}\ }\textbf {\bibinfo {volume}
  {253}},\ \bibinfo {pages} {349} (\bibinfo {year} {1959})}\BibitemShut
  {NoStop}%
\bibitem [{\citenamefont {{Harvey}}\ and\ \citenamefont
  {{Forgham}}(1984)}]{Harvey-Forgham:1984}%
  \BibitemOpen
  \bibfield  {author} {\bibinfo {author} {\bibfnamefont {J.~E.}\ \bibnamefont
  {{Harvey}}}\ and\ \bibinfo {author} {\bibfnamefont {J.~L.}\ \bibnamefont
  {{Forgham}}},\ }\bibfield  {title} {\bibinfo {title} {The spot of arago: New
  relevance for an old phenomenon},\ }\href@noop {} {\bibfield  {journal}
  {\bibinfo  {journal} {Am. J. Phys.}\ }\textbf {\bibinfo {volume} {52}},\
  \bibinfo {pages} {243} (\bibinfo {year} {1984})}\BibitemShut {NoStop}%
\bibitem [{\citenamefont {Turyshev}\ and\ \citenamefont
  {Toth}(2023)}]{Turyshev-Toth:2023}%
  \BibitemOpen
  \bibfield  {author} {\bibinfo {author} {\bibfnamefont {S.~G.}\ \bibnamefont
  {Turyshev}}\ and\ \bibinfo {author} {\bibfnamefont {V.~T.}\ \bibnamefont
  {Toth}},\ }\bibfield  {title} {\bibinfo {title} {{Evolving morphology of
  resolved stellar Einstein rings}},\ }\href
  {https://doi.org/10.3847/1538-4357/acaf4f} {\bibfield  {journal} {\bibinfo
  {journal} {ApJ}\ }\textbf {\bibinfo {volume} {944}},\ \bibinfo {pages} {25}
  (\bibinfo {year} {2023})},\ \Eprint {https://arxiv.org/abs/{arXiv:2209.09534
  [astro-ph.IM]}} {{arXiv:2209.09534 [astro-ph.IM]}} \BibitemShut {NoStop}%
\bibitem [{\citenamefont {{Abramowitz}}\ and\ \citenamefont
  {{Stegun}}(1965)}]{Abramovitz-Stegun:1965}%
  \BibitemOpen
  \bibfield  {author} {\bibinfo {author} {\bibfnamefont {M.}~\bibnamefont
  {{Abramowitz}}}\ and\ \bibinfo {author} {\bibfnamefont {I.~A.}\ \bibnamefont
  {{Stegun}}},\ }\href@noop {} {\emph {\bibinfo {title} {Handbook of
  Mathematical Functions: with Formulas, Graphs, and Mathematical Tables.}}}\
  (\bibinfo  {publisher} {Dover Publications, New York; revised edition},\
  \bibinfo {year} {1965})\BibitemShut {NoStop}%
\bibitem [{\citenamefont {Baumbach}(1937)}]{Baumbach:1937}%
  \BibitemOpen
  \bibfield  {author} {\bibinfo {author} {\bibfnamefont {S.}~\bibnamefont
  {Baumbach}},\ }\bibfield  {title} {\bibinfo {title} {Strahlung, ergiebigkeit
  und elektronendichte der sonnenkorona},\ }\href@noop {} {\bibfield  {journal}
  {\bibinfo  {journal} {Astronomische Nachrichten}\ }\textbf {\bibinfo {volume}
  {263}},\ \bibinfo {pages} {121} (\bibinfo {year} {1937})}\BibitemShut
  {NoStop}%
\bibitem [{\citenamefont {{Golub}}\ and\ \citenamefont
  {{Pasachoff}}(2009)}]{Golub-Pasachoff-book:2017}%
  \BibitemOpen
  \bibfield  {author} {\bibinfo {author} {\bibfnamefont {L.}~\bibnamefont
  {{Golub}}}\ and\ \bibinfo {author} {\bibfnamefont {J.~M.}\ \bibnamefont
  {{Pasachoff}}},\ }\href@noop {} {\emph {\bibinfo {title} {The Solar
  Corona}}}\ (\bibinfo  {publisher} {2-nd edition. Cambridge University
  Press},\ \bibinfo {address} {Cambridge, England},\ \bibinfo {year}
  {2009})\BibitemShut {NoStop}%
\bibitem [{\citenamefont {{November}}\ and\ \citenamefont
  {{Koutchmy}}(1996)}]{November:1996}%
  \BibitemOpen
  \bibfield  {author} {\bibinfo {author} {\bibfnamefont {L.~J.}\ \bibnamefont
  {{November}}}\ and\ \bibinfo {author} {\bibfnamefont {S.}~\bibnamefont
  {{Koutchmy}}},\ }\bibfield  {title} {\bibinfo {title} {White-light coronal
  dark threads and density fine structure},\ }\href
  {https://doi.org/10.1086/177528} {\bibfield  {journal} {\bibinfo  {journal}
  {ApJ}\ }\textbf {\bibinfo {volume} {466}},\ \bibinfo {pages} {512} (\bibinfo
  {year} {1996})}\BibitemShut {NoStop}%
\bibitem [{\citenamefont {Turyshev}\ and\ \citenamefont
  {Toth}(2019)}]{Turyshev-Toth:2019}%
  \BibitemOpen
  \bibfield  {author} {\bibinfo {author} {\bibfnamefont {S.~G.}\ \bibnamefont
  {Turyshev}}\ and\ \bibinfo {author} {\bibfnamefont {V.~T.}\ \bibnamefont
  {Toth}},\ }\bibfield  {title} {\bibinfo {title} {Diffraction of light by the
  gravitational field of the sun and the solar corona},\ }\href@noop {}
  {\bibfield  {journal} {\bibinfo  {journal} {Phys. Rev. D}\ }\textbf {\bibinfo
  {volume} {99}},\ \bibinfo {pages} {024044} (\bibinfo {year} {2019})},\
  \Eprint {https://arxiv.org/abs/arXiv:1810.06627 [gr-qc]} {arXiv:1810.06627
  [gr-qc]} \BibitemShut {NoStop}%
\bibitem [{\citenamefont {{Turyshev}}\ and\ \citenamefont
  {{Toth}}(2019)}]{Turyshev-Toth:2019-corona}%
  \BibitemOpen
  \bibfield  {author} {\bibinfo {author} {\bibfnamefont {S.~G.}\ \bibnamefont
  {{Turyshev}}}\ and\ \bibinfo {author} {\bibfnamefont {V.~T.}\ \bibnamefont
  {{Toth}}},\ }\bibfield  {title} {\bibinfo {title} {{Optical properties of the
  solar gravitational lens in the presence of the solar corona}},\ }\href
  {https://doi.org/10.1140/epjp/i2019-12426-4} {\bibfield  {journal} {\bibinfo
  {journal} {Eur. Phys. J. Plus}\ }\textbf {\bibinfo {volume} {134}},\ \bibinfo
  {eid} {63} (\bibinfo {year} {2019})},\ \Eprint
  {https://arxiv.org/abs/arXiv:1811.06515 [gr-qc]} {arXiv:1811.06515 [gr-qc]}
  \BibitemShut {NoStop}%
\bibitem [{\citenamefont {{Hippke}}(2018)}]{Hippke:2018}%
  \BibitemOpen
  \bibfield  {author} {\bibinfo {author} {\bibfnamefont {M.}~\bibnamefont
  {{Hippke}}},\ }\bibfield  {title} {\bibinfo {title} {{Interstellar
  communication. II. Application to the solar gravitational lens}},\ }\href
  {https://doi.org/10.1016/j.actaastro.2017.10.022} {\bibfield  {journal}
  {\bibinfo  {journal} {Acta Astronaut.}\ }\textbf {\bibinfo {volume} {142}},\
  \bibinfo {pages} {64} (\bibinfo {year} {2018})}\BibitemShut {NoStop}%
\bibitem [{\citenamefont {{Marcy}}\ \emph {et~al.}(2022)\citenamefont
  {{Marcy}}, \citenamefont {{Tellis}},\ and\ \citenamefont
  {{Wishnow}}}]{Marcy:2022}%
  \BibitemOpen
  \bibfield  {author} {\bibinfo {author} {\bibfnamefont {G.~W.}\ \bibnamefont
  {{Marcy}}}, \bibinfo {author} {\bibfnamefont {N.~K.}\ \bibnamefont
  {{Tellis}}},\ and\ \bibinfo {author} {\bibfnamefont {E.~H.}\ \bibnamefont
  {{Wishnow}}},\ }\bibfield  {title} {\bibinfo {title} {{Laser communication
  with Proxima and Alpha Centauri using the solar gravitational lens}},\ }\href
  {https://doi.org/10.1093/mnras/stab3074} {\bibfield  {journal} {\bibinfo
  {journal} {MNRAS}\ }\textbf {\bibinfo {volume} {509}},\ \bibinfo {pages}
  {3798} (\bibinfo {year} {2022})}\BibitemShut {NoStop}%
\bibitem [{\citenamefont {{Helvajian}}\ \emph {et~al.}(2023)\citenamefont
  {{Helvajian}}, \citenamefont {{Rosenthal}}, \citenamefont {{Poklemba}},
  \citenamefont {{Battista}}, \citenamefont {{DiPrinzio}}, \citenamefont
  {{Neff}}, \citenamefont {{McVey}}, \citenamefont {{Toth}},\ and\
  \citenamefont {{Turyshev}}}]{Helvajian-etal:2022}%
  \BibitemOpen
  \bibfield  {author} {\bibinfo {author} {\bibfnamefont {H.}~\bibnamefont
  {{Helvajian}}}, \bibinfo {author} {\bibfnamefont {A.}~\bibnamefont
  {{Rosenthal}}}, \bibinfo {author} {\bibfnamefont {J.}~\bibnamefont
  {{Poklemba}}}, \bibinfo {author} {\bibfnamefont {T.~A.}\ \bibnamefont
  {{Battista}}}, \bibinfo {author} {\bibfnamefont {M.~D.}\ \bibnamefont
  {{DiPrinzio}}}, \bibinfo {author} {\bibfnamefont {J.~M.}\ \bibnamefont
  {{Neff}}}, \bibinfo {author} {\bibfnamefont {J.~P.}\ \bibnamefont {{McVey}}},
  \bibinfo {author} {\bibfnamefont {V.~T.}\ \bibnamefont {{Toth}}},\ and\
  \bibinfo {author} {\bibfnamefont {S.~G.}\ \bibnamefont {{Turyshev}}},\
  }\bibfield  {title} {\bibinfo {title} {{Mission Architecture to Reach and
  Operate at the Focal Region of the Solar Gravitational Lens}},\ }\href
  {https://doi.org/10.2514/1.A35493} {\bibfield  {journal} {\bibinfo  {journal}
  {J. Spacecraft \& Rockets}\ }\textbf {\bibinfo {volume} {60}},\ \bibinfo
  {pages} {829} (\bibinfo {year} {2023})},\ \Eprint
  {https://arxiv.org/abs/arXiv:2207.03005 [astro-ph.IM]} {arXiv:2207.03005
  [astro-ph.IM]} \BibitemShut {NoStop}%
\bibitem [{\citenamefont {{Turyshev}}\ \emph {et~al.}(2023)\citenamefont
  {{Turyshev}}, \citenamefont {{Garber}}, \citenamefont {{Friedman}},
  \citenamefont {{Hein}}, \citenamefont {{Barnes}}, \citenamefont {{Batygin}},
  \citenamefont {{Brown}}, \citenamefont {{Cronin}}, \citenamefont {{Davoyan}},
  \citenamefont {{Dubill}}, \citenamefont {{Eubanks}}, \citenamefont
  {{Gibson}}, \citenamefont {{Hassler}}, \citenamefont {{Izenberg}},
  \citenamefont {{Kervella}}, \citenamefont {{Mauskopf}}, \citenamefont
  {{Murphy}}, \citenamefont {{Nutter}}, \citenamefont {{Porco}}, \citenamefont
  {{Riccobono}}, \citenamefont {{Schalkwyk}}, \citenamefont {{Stevenson}},
  \citenamefont {{Sykes}}, \citenamefont {{Sultana}}, \citenamefont {{Toth}},
  \citenamefont {{Velli}},\ and\ \citenamefont
  {{Worden}}}]{Turyshev-etal:2023}%
  \BibitemOpen
  \bibfield  {author} {\bibinfo {author} {\bibfnamefont {S.~G.}\ \bibnamefont
  {{Turyshev}}}, \bibinfo {author} {\bibfnamefont {D.}~\bibnamefont
  {{Garber}}}, \bibinfo {author} {\bibfnamefont {L.~D.}\ \bibnamefont
  {{Friedman}}}, \bibinfo {author} {\bibfnamefont {A.~M.}\ \bibnamefont
  {{Hein}}}, \bibinfo {author} {\bibfnamefont {N.}~\bibnamefont {{Barnes}}},
  \bibinfo {author} {\bibfnamefont {K.}~\bibnamefont {{Batygin}}}, \bibinfo
  {author} {\bibfnamefont {M.~E.}\ \bibnamefont {{Brown}}}, \bibinfo {author}
  {\bibfnamefont {L.}~\bibnamefont {{Cronin}}}, \bibinfo {author}
  {\bibfnamefont {A.~R.}\ \bibnamefont {{Davoyan}}}, \bibinfo {author}
  {\bibfnamefont {A.}~\bibnamefont {{Dubill}}}, \bibinfo {author}
  {\bibfnamefont {T.~M.}\ \bibnamefont {{Eubanks}}}, \bibinfo {author}
  {\bibfnamefont {S.}~\bibnamefont {{Gibson}}}, \bibinfo {author}
  {\bibfnamefont {D.~M.}\ \bibnamefont {{Hassler}}}, \bibinfo {author}
  {\bibfnamefont {N.~R.}\ \bibnamefont {{Izenberg}}}, \bibinfo {author}
  {\bibfnamefont {P.}~\bibnamefont {{Kervella}}}, \bibinfo {author}
  {\bibfnamefont {P.~D.}\ \bibnamefont {{Mauskopf}}}, \bibinfo {author}
  {\bibfnamefont {N.}~\bibnamefont {{Murphy}}}, \bibinfo {author}
  {\bibfnamefont {A.}~\bibnamefont {{Nutter}}}, \bibinfo {author}
  {\bibfnamefont {C.}~\bibnamefont {{Porco}}}, \bibinfo {author} {\bibfnamefont
  {D.}~\bibnamefont {{Riccobono}}}, \bibinfo {author} {\bibfnamefont
  {J.}~\bibnamefont {{Schalkwyk}}}, \bibinfo {author} {\bibfnamefont {K.~B.}\
  \bibnamefont {{Stevenson}}}, \bibinfo {author} {\bibfnamefont {M.~V.}\
  \bibnamefont {{Sykes}}}, \bibinfo {author} {\bibfnamefont {M.}~\bibnamefont
  {{Sultana}}}, \bibinfo {author} {\bibfnamefont {V.~T.}\ \bibnamefont
  {{Toth}}}, \bibinfo {author} {\bibfnamefont {M.}~\bibnamefont {{Velli}}},\
  and\ \bibinfo {author} {\bibfnamefont {S.~P.}\ \bibnamefont {{Worden}}},\
  }\bibfield  {title} {\bibinfo {title} {{Science opportunities with solar
  sailing smallsats}},\ }\href {https://doi.org/10.1016/j.pss.2023.105744}
  {\bibfield  {journal} {\bibinfo  {journal} {Planetary \& Space Sci.}\
  }\textbf {\bibinfo {volume} {235}},\ \bibinfo {eid} {105744} (\bibinfo {year}
  {2023})},\ \Eprint {https://arxiv.org/abs/arXiv:2303.14917 [astro-ph.EP]}
  {arXiv:2303.14917 [astro-ph.EP]} \BibitemShut {NoStop}%
\bibitem [{\citenamefont {{Turyshev}}\ and\ \citenamefont
  {{Toth}}(2021)}]{Turyshev-Toth:2021-caustics}%
  \BibitemOpen
  \bibfield  {author} {\bibinfo {author} {\bibfnamefont {S.~G.}\ \bibnamefont
  {{Turyshev}}}\ and\ \bibinfo {author} {\bibfnamefont {V.~T.}\ \bibnamefont
  {{Toth}}},\ }\bibfield  {title} {\bibinfo {title} {Optical properties of an
  extended gravitational lens},\ }\href@noop {} {\bibfield  {journal} {\bibinfo
   {journal} {Phys. Rev. D}\ }\textbf {\bibinfo {volume} {104}},\ \bibinfo
  {pages} {024019} (\bibinfo {year} {2021})},\ \bibinfo {note}
  {arXiv:2103.06955 [gr-qc]}\BibitemShut {NoStop}%
\bibitem [{\citenamefont {Turyshev}\ and\ \citenamefont
  {Toth}(2021)}]{Turyshev-Toth:2021-quartic}%
  \BibitemOpen
  \bibfield  {author} {\bibinfo {author} {\bibfnamefont {S.~G.}\ \bibnamefont
  {Turyshev}}\ and\ \bibinfo {author} {\bibfnamefont {V.~T.}\ \bibnamefont
  {Toth}},\ }\bibfield  {title} {\bibinfo {title} {Wave-optical study of the
  einstein cross formed by a quadrupole gravitational lens},\ }\href@noop {}
  {\bibfield  {journal} {\bibinfo  {journal} {Phys. Rev. D}\ }\textbf {\bibinfo
  {volume} {104}},\ \bibinfo {pages} {124033} (\bibinfo {year} {2021})},\
  \bibinfo {note} {arXiv:2105.07295 [gr-qc]}\BibitemShut {NoStop}%
\bibitem [{\citenamefont {{Turyshev}}\ and\ \citenamefont
  {{Toth}}(2023)}]{Turyshev-Toth:2023-faint}%
  \BibitemOpen
  \bibfield  {author} {\bibinfo {author} {\bibfnamefont {S.~G.}\ \bibnamefont
  {{Turyshev}}}\ and\ \bibinfo {author} {\bibfnamefont {V.~T.}\ \bibnamefont
  {{Toth}}},\ }\bibfield  {title} {\bibinfo {title} {{Imaging faint sources
  with the extended solar gravitational lens}},\ }\href
  {https://doi.org/10.1103/PhysRevD.107.104063} {\bibfield  {journal} {\bibinfo
   {journal} {Phys. Rev. D}\ }\textbf {\bibinfo {volume} {107}},\ \bibinfo
  {eid} {104063} (\bibinfo {year} {2023})},\ \Eprint
  {https://arxiv.org/abs/arXiv:2301.07495 [astro-ph.IM]} {arXiv:2301.07495
  [astro-ph.IM]} \BibitemShut {NoStop}%
\bibitem [{\citenamefont {{Turyshev}}\ and\ \citenamefont
  {{Toth}}(2020{\natexlab{b}})}]{Turyshev-Toth:2020-image}%
  \BibitemOpen
  \bibfield  {author} {\bibinfo {author} {\bibfnamefont {S.~G.}\ \bibnamefont
  {{Turyshev}}}\ and\ \bibinfo {author} {\bibfnamefont {V.~T.}\ \bibnamefont
  {{Toth}}},\ }\bibfield  {title} {\bibinfo {title} {Image formation process
  with the solar gravitational lens},\ }\href
  {https://doi.org/10.1103/PhysRevD.101.044048} {\bibfield  {journal} {\bibinfo
   {journal} {Phys. Rev. D}\ }\textbf {\bibinfo {volume} {101}},\ \bibinfo
  {pages} {044048} (\bibinfo {year} {2020}{\natexlab{b}})},\ \bibinfo {note}
  {arXiv:1911.03260 [gr-qc]}\BibitemShut {NoStop}%
\bibitem [{\citenamefont {{Feynman}}\ and\ \citenamefont
  {{Hibbs}}(1965)}]{Feynman-Hibbs:1991}%
  \BibitemOpen
  \bibfield  {author} {\bibinfo {author} {\bibfnamefont {R.}~\bibnamefont
  {{Feynman}}}\ and\ \bibinfo {author} {\bibfnamefont {A.}~\bibnamefont
  {{Hibbs}}},\ }\href@noop {} {\emph {\bibinfo {title} {Quantum Mechanics and
  Path Integrals}}}\ (\bibinfo  {publisher} {McGraw-Hill},\ \bibinfo {address}
  {New York},\ \bibinfo {year} {1965})\BibitemShut {NoStop}%
\end{thebibliography}

%

\end{document}